\documentclass[11pt]{article}
\usepackage{amssymb,amsmath,amsthm}
\usepackage{graphicx}
\usepackage{hyperref}

\usepackage{booktabs,multirow}
\usepackage{caption}
\usepackage{fancyvrb}
\usepackage{amsfonts}
\usepackage{subfloat}
\usepackage{wrapfig}
\usepackage{upgreek}
\usepackage[square,sort,numbers]{natbib}
\usepackage{footnote}
\usepackage{lipsum}
\usepackage{overpic}
\usepackage{subfig}
\usepackage{eqnarray}
\usepackage{float}
\usepackage{colortbl}

\usepackage{rotating}
\usepackage{pifont} 
\usepackage{commath}
\usepackage{xcolor}
\usepackage{multicol}
\usepackage{enumitem}
\setlist{itemsep=0.2pt}
\usepackage{textcomp} 
\usepackage{comment}

\theoremstyle{definition}

\usepackage{todonotes}
\usepackage{fancyhdr}
\begin{document}
\vspace{-1cm}


\date{28 February 2020}
\title{146$^{th}$ European Study Group with Industry/\\Co-creation event with society\\
\vspace{0.5cm} Breaking barriers for women in science}
\maketitle
\noindent
\begin{center}
\textbf{Report authors}\\
Demetris Avraam, Department of Business and Public Administration, \\University of Cyprus, CY
\\ 
Vasiliki Bitsouni, SciCo Cyprus, CY \& The Roslin Institute, \\University of Edinburgh, UK \\ 
Nikoleta E. Glynatsi, School of Mathematics,
Cardiff University, UK \\
Katerina Kaouri, School of Mathematics,
Cardiff University, UK \\ 
Alessandra Micheletti, Department of Environmental Science and Policy,
Universit\'a degli Studi di Milano, IT \\ 
M. Ros\'ario Oliveira, Mathematics Department and CEMAT, Instituto Superior
T\'ecnico,
Universidade de Lisboa,  PT 
\\
Margarita Zachariou, Bioinformatics Group,
Cyprus Institute of Neurology \& Genetics, CY 
\end{center}
\noindent
\begin{center}
\textbf{Other team participants}\\
Natalie Christopher, European University Cyprus, CY\\
Poul Hjorth, Technical University of Denmark, DK\\
Timos Papadopoulos, Public Health England, UK \\
Ioulia Televantou, University of Cyprus, CY \\
Kyriacos Vitalis, University of Cyprus, CY\\
\end{center}

\noindent
\begin{center}
\textbf{Problem proposer:}\\
Anna Koukkides-Procopiou, AIPFE Cyprus-Women of Europe, CY\\
\end{center}

\noindent
\begin{center}
\textbf{Academic coordinators:}\\
Katerina Kaouri, Margarita Zachariou
\end{center}
\newpage
\begin{abstract}
This report summarises the work and results produced at the 146th European Study
Group with Industry/co-creation event with society on the challenge \textit{Breaking barriers for women in Science}. The aim of this challenge, proposed by the Cyprus-based non-profit AIPFE Cyprus-Women of Europe, was to quantify the barriers that women face in science so that eventually policy changes can take place in Cyprus and elsewhere. Two distinct but related challenges were considered. The first challenge was to quantify the wage gap between men and women in 28 European countries. In this connection, we analysed Eurostat data and developed a mathematical model quantifying how probable it is for countries to decrease their wage gap. Secondly, we analysed data provided by the University of Cyprus and determined the percentage of women and men in STEM (Science, Technology, Engineering and Mathematics) departments as they move up the academic ladder, starting from the undergraduate level.
Studying the latter challenge is a first step in studying the wage gap in all Cypriot universities and in other universities abroad. This work was supported financially by the EU project SciShops.eu and several other organisations.
\

{\sc Keywords:} data analysis; stochastic processes; generalized linear models.

\end{abstract}


\section{Introduction}
This report summarises the results from the week-long teamwork during the 146th European Study
Group with Industry (ESGI146)/co-creation event with society, which was held in Cyprus, from 3-7 December 2018. The challenge was to quantify the barriers that women in science face, in Cyprus and abroad. The work was undertaken in collaboration with the Cypriot-based non-profit AIPFE Cyprus-Women of Europe. Prior to ESGI146 a two-hour co-creation workshop (funded by SciShops.eu) was
run on the 29th of October 2018, at the Cyprus Institute of Neurology and Genetics where 38 stakeholders from academia, businesses, civil society and policy makers participated and discussed key challenges. Subsequently, the challenges were reviewed and two were chosen for the ESGI146 workshop.  The first challenge was to investigate the wage gap between men and women and create a mathematical model quantifying how probable it is for a country to reduce its wage gap. The second challenge was to analyse data from the University of Cyprus in order to quantify the progression of women on the academic ladder, in STEM departments. Note that UCY was chosen as a pilot, as the oldest university on the island but the aim is to eventually conduct the same analysis in all other universities in Cyprus and other countries.

This report is structured as follows:
\begin{itemize}
  \item\textbf{Section~\ref{section:description}} contains a description of the
  challenges presented at ESGI146.
  \item\textbf{Section~\ref{Sec:ModelMC}} describes the methodology used to
  develop a stochastic model to study the wage gap between men and women.
  \item\textbf{Section~\ref{section:cyprus_data}} presents the analysis of data provided by the University of Cyprus; some possible next steps are identified.
\end{itemize}

\section{Description of the challenges}\label{section:description}
The two challenges was as follows:

(1) Describe the evolution of a gender wage gap indicator in the EU. A mathematical model was developed to explore which factors affect the wage gap between the two genders in 28 European countries. The factors taken into account in the modelling were \textit{Economics, Health, Education, and Political Participation}.
(2) Form a better understanding of the gender gap in Cypriot academia, in the STEM departments. 

The data for this study were provided by
\begin{itemize}
   \item Eurostat \cite{eurostat}
    \item The University of Cyprus (UCY)
    \item The Statistical Service of Cyprus
\end{itemize}

\section{Mathematical models and data analysis}

This section describes the mathematical modelling and data analysis conducted in order to tackle the challenges described above. It is divided into two subsections.
\subsection{A gender wage gap indicator}\label{Sec:ModelMC}

We consider a gender wage gap indicator and we assume that we have recorded its evolution in time. We also assume to have data on the evolution of other socioeconomic indicators during the same time period. We will
investigate the dependence of the gender wage gap indicator on the other socioeconomic indicators. Since gender wage gap data have only recently been collected, and
publicly available indicators are often computed and released only once per year for
each country or region of study, it is difficult to recover long histories
of such indicators. Thus, our mathematical model must depend only on a small number of unknown parameters, which can be estimated even from relatively small datasets.

We consider here the evolution of the gender wage gap indicator in
28 different European countries, during 9 years, from 2009 to 2016. The data have been downloaded from Eurostat \cite{eurostat}. The gender wage gap
$X$ is defined as:

$$X\colon=\frac{mean\  daily\  wage\ of\ men - mean\ daily\ wage\ of\
women}{mean\  daily\  wage\ of\ men}.$$

We will test the existence of any significant dependence of $X$ on
the following four indicators:
\begin{itemize}
\item Economics = females/males ratio, employment rate of highly educated
individuals (tertiary education level)
\item Health = females/males ratio, life expectancy 
\item Education = females/males ratio of highly educated individuals (tertiary
educ.)
\item Political involvement =females/male ratio, presence in the country Parliament, 
\end{itemize}
We decided to use these indicators firstly due to the availability of the yearly data on the Eurostat website for the same countries and years considered for the wage gap,
and secondly because their contribution to the formation of four similar indicators were used by the World Economic Forum to build a Global Gender Gap Index in their 2017 report \cite{WEF}.

\subsubsection{A Markov Chain model for the wage gap indicator}\label{Sec:MM}
We describe the evolution of the wage gap $X$ as a Markov Chain \cite{Norris}, as follows.

Let $X_t$ be the wage gap of a given country at year $t$. We discretize the
state space of $X_t$ so that,

$$
X_t\colon\Omega\to S=\{s_1,\dots,s_m\}
$$

where $s_i=i-th$ level of wage gap and $m$ is a finite positive integer. We use
this discretization and consider a small  number, $m$, of different levels
in the wage gap. This approach both reduces the number of unknown parameters to be estimated
from the data remains small and enables the  identification of these parameters  even for
relatively small datasets.

We assume that the Markov property holds, i.e., the evolution
of the wage gap in a country in the subsequent year depends only on the present value of its wage gap, and
not on the entire history of the process:
\begin{eqnarray*}
&&P(X_{t+1}=s_j|X_t=s_i,X_{t-1}=s_{l_1},X_{t-2}=s_{l_2},\dots , X_0=s_{l_t})\\
&=&P(X_{t+1}=s_j|X_t=s_i)=\colon p_{ij}(t).
\end{eqnarray*}
Here, $p_{ij}(t)$ is the probability that any country changes from level $s_i$ to
level $s_j$ in one time step. Initially, we explore whether any significant relationship
exists between the transition probabilities and the other indicators (either all or a subset)
$$p_{ij}(t)=f(economical(t),political(t),health(t),educational(t))$$.

We assume to have a sample $\{X_t^1\}_{t=1}^k, \dots, \{X_t^n\}_{t=1}^k$ of $n$
i.i.d. processes observed on the same time span $t=1,\dots,k$ (i.e., in our
case $n=28$ corresponding to EU-28, over the time period
$t=2009,\dots , 2016$). Then the transition probabilities $p_{ij}$ can be
estimated through:

\ 

\noindent $\hat{p}_{ij}(t)=$proportion of countries that passed from $s_i$ to
$s_j$ from time $t-1$ to time $t$.

\ 

We discretized the wage gap into 3 levels, i.e. $m=3$, by dividing the
distribution of the wage gaps into 3 equal parts, considered independently of time. Thus "level 1
countries" are the best (smaller wage gap) while "level 3 countries" are the
worst (largest wage gap). We then estimated the transition matrices using the
proportions of countries which are changing level, year by year, and we obtained:

$$
 P(2009\to 2010)=\begin{bmatrix}
0.9 & 0.1 & 0\\
0.15 & 0.85 & 0\\
0 & 0.09 & 0.91
\end{bmatrix}
$$
$$
P(2010\to 2011)=\begin{bmatrix}
1 & 0 & 0\\
0 & 0.875 & 0.125\\
0 & 0.1 & 0.9
\end{bmatrix}
$$
$$
\vdots
$$
$$
P(2015\to 2016)=\begin{bmatrix}
1 & 0 & 0\\
0 & 1 & 0 \\
0 & 0.14 & 0.86
\end{bmatrix}.
$$
We also computed the \emph{overall transition matrix} (computed without taking
time into account), obtaining the following estimate
\begin{equation}
P_{(2009\to 2016)}=\begin{bmatrix}
0.925 & 0.075 & 0\\
0.08 & 0.88 & 0.04\\
0 & 0.12 & 0.88
\end{bmatrix}. \label{overallP}
\end{equation}
From these results we observe that
\begin{itemize}
\item There is a tendency of countries to remain at wage gap level,
\item ``jumps" of two levels in one year are highly unlikely,
\item level 2 countries (medium) have about the same (small) probability to
improve or to get worse.
\end{itemize}
In the next section We will take into account these observations in order to fit a suitable model which
relates the transition probabilities with the four indicators
\textit{Economical, Health, Educational, Political} (which we will call covariates from now on).

\subsubsection{A multinomial logistic regression model for the evolution of transition probabilities}\label{Sec:Log_Reg}

The relationship between the transition probabilities and the other indicators
can then be studied using a \emph{multinomial logistic regression model} (also
called \emph{polytomous logistic regression model})
\cite{Agresti,Castilla-etal,Engel88}. These models, for $m$ possible outcomes
(i.e. wage gap levels), run $m-1$ independent binary
logistic regression models, in which one outcome is chosen as a ``baseline" and
then the other $m-1$ outcomes are separately regressed against the baseline
outcome. We consider the transition matrix at time $t$:

$$
P(t)=\begin{bmatrix}
    p_{11}(t) & p_{12}(t) & p_{13}(t) & \dots  & p_{1m}(t) \\
    p_{21}(t) & p_{22}(t) & p_{23}(t) & \dots  & p_{2m}(t) \\
    \vdots & \vdots & \vdots & \ddots & \vdots \\
    p_{m1}(t) & p_{m2}(t) & p_{m3}(t) & \dots  & p_{mm}(t)
\end{bmatrix}
$$

In our case, since we have only three wage gap levels, $m=3$ and $P(t)$ is a $3\times 3$ matrix. Each row of $P(t)$ forms a discrete probability distribution ($\sum_{j=1}^m p_{ij}(t)=1$) and we will study a different multinomial
regression model for each row $i$, using as baseline outcome the $i$-th wage gap
level. Denoting by $\underline{Z}(t)$ the vector of the four covariates, at time $t$, for a given
country, the model is stated as the following set of equations:
\begin{eqnarray}
\ln \frac{p_{i1}(t)}{p_{ii}(t)}&=&\underline{\beta}_1\cdot \underline{Z}(t)
\nonumber \\
\ln \frac{p_{i2}(t)}{p_{ii}(t)}&=&\underline{\beta}_2\cdot
\underline{Z}(t)\nonumber\\
&&\dots \nonumber\\
\ln\frac{p_{i(i-1)}(t)}{p_{ii}(t)}&=&\underline{\beta}_{i-1}\cdot
\underline{Z}(t) \label{multimodel}\\ 
\ln\frac{p_{i(i+1)}(t)}{p_{ii}(t)}&=&\underline{\beta}_{i+1}\cdot
\underline{Z}(t)\nonumber\\ 
&&\dots \nonumber\\
\ln\frac{p_{im}(t)}{p_{ii}(t)}&=&\underline{\beta}_{m}\cdot
\underline{Z}(t)\nonumber
\end{eqnarray}

for $i=1,\dots ,m$. In this model, the components of the coefficients
$\underline{\beta}_k$ should be interpreted as the contribution of the single
covariates (components of $\underline{Z}$) to the relative variation of the
probabilities of changing state in one time step, with respect to the
probability to remain in the same state (also called \emph{odds ratio}). A
positive sign in a coefficient means that an increase of the corresponding
covariate contributes to increase the (relative) probability of changing level,
while a negative sign means that an increase of the corresponding covariate
contributes to decrease the (relative) probability of changing level. 

Note that the choice of using as baseline outcome the state from which the process is
evolving, is due to our previous observation that the wage gap has a tendency to
remain fixed at the same level for most countries. With this choice we highlight
the contribution of the covariates to move the wage gap from a static situation.

We also assume that the coefficients $\underline{\beta}_k$ are constant in time,
which corresponds to the relationship between the covariates
$\underline{Z}$ and the odds ratios remaining unchanged over time. This
assumption allows us to merge data from different years to estimate the
parameters, enriching thus the sample and increasing the reliability of the
estimates\footnote{The unknown parameters in each vector $\underline{\beta}_k$ are typically
jointly estimated by iterative numerical methods to maximize the likelihood, or
by maximum a posterior (MAP) estimation, which is an extension of maximum
likelihood using regularization of the weights to prevent pathological
solutions. Alternatively, minimum density power divergence estimators have been
proposed as more robust estimators \cite{Castilla-etal}. We estimated the
parameters using the {\it multinom} function in the R package {\it nnet}, which
fits multinomial log-linear models via neural networks. We also computed
Wald tests to evaluate the significance of the coefficients, which are approximate,
since are based on the assumption of normality of the residuals, which in nonlinear models is not satisfied, in general.}

\subsubsection{Numerical results of the multinomial regression model}

We applied model (\ref{multimodel}) to our data and estimated the parameters.
Note that we discretized the wage gap into three levels (resulting in a 
 $3\times 3$ transition matrix) and we then performed separate estimations of
the $\underline{\beta}_k$ coefficients for the three rows of the matrix. Let us
now analyse the results row by row.

\ 

\noindent{\bf Row 1 (Level 1 countries)}. The three elements $p_{11}(t),p_{12}(t),p_{13}(t)$ of the
first row of the transition matrix represent the probabilities that a country
which is at level 1 of wage gap (the best level) at time $t$ passes to level
1,2,3, respectively, at time $t+1$. Actually, observing the estimated overall
transition matrix (\ref{overallP}) and also the single estimated transition
matrices, reported in the previous section, we fix $p_{13}(t)=0$ for all
considered years $t$. Thus in model (\ref{multimodel}) only the two
probabilities $p_{11}(t)$ and $p_{12}(t)$ survive, and the model reduces to the
usual logistic regression model

$$
\ln \frac{p_{12}(t)}{p_{11}(t)}=\underline{\beta}\cdot \underline{Z}(t)
$$

where we have to estimate only one vector $\underline{\beta}$ of coefficients,
whose elements tell us the contribution of each covariate in the (relative)
probability of {\bf increasing} the wage gap from level 1 to level 2. We
obtained the results reported in Table \ref{row1}.

\begin{table}[h]
    \begin{center}
    \begin{tabular}{lrrr}
        \toprule
        {\bf Coefficients} &  {\bf Values} &  {\bf Std. Err.} & {\bf Wald test pvalues}\\ \midrule
        Intercept          &  -40.267761   & 24.510129        & 0.10040304\\
        Economics          &     7.165891  & 3.734866         & 0.05502896 \\
        Education          &  7.165891     & 3.734866         & 0.05502896\\
        Health             &  13.763583    & 18.791091        & 0.46389239\\
        Politics           &   5.173787    &  5.006702        & 0.30142973 \\
        \bottomrule
    \end{tabular}
    \end{center}
    \caption{Estimates of the $\underline{\beta}$ coefficient for {\bf row 1} of the transition matrix.
    The Wald test is testing the null hypothesis that the corresponding coefficient is equal to zero.
    Using a significance level of $90\%$, only the coefficients having a p-value below 0.1 are significant
    (i.e. the null hypothesis is rejected)}\label{row1}
\end{table}
We observe that only the coefficients of the Economics and Education indicators
are significant and have a positive sign. This means that, for countries which
start from level 1 (the best level), an improvement in the
women/men employment rate and in the women/men higher education is connected
with a small increase of the probability that the country worsens its situation,
passing to level 2.

\ 

\noindent{\bf Row 2 (Level 2 countries).}
The three elements $p_{21}(t),p_{22}(t),p_{23}(t)$ of the first row of the
transition matrix represent the probabilities that a country in level 2 (medium level) at time $t$ passes to level 1, 2, 3,
respectively, at time $t+1$. Then $p_{21}(t)$ represents the probability to
reduce the wage gap, and improve the situation, while $p_{23}$ represents the
probability to increase the wage gap and thus worsen the situation. We observe
from the estimated overall transition matrix (\ref{overallP}) that all the
elements of the second row are different from zero, and therefore, in this case the model
(\ref{multimodel}) reads as follows:
\begin{eqnarray*}
\ln \frac{p_{21}(t)}{p_{22}(t)}&=&\underline{\beta}_1\cdot \underline{Z}(t)  \\
\ln \frac{p_{23}(t)}{p_{22}(t)}&=&\underline{\beta}_3\cdot \underline{Z}(t) \\
\end{eqnarray*}
The results of the estimates of the coefficients $\underline{\beta}_1,
\underline{\beta}_3$ are reported in Table \ref{row2}.

\begin{table}[h]
    \begin{center}
    \begin{tabular}{lrrrrr}
    \multicolumn{6}{c}{\bf Coefficients}\\
    \toprule
    \ & (Intercept) & Economics & Education & Health  & Political \\ \midrule
    $\underline{\beta}_1$  &  1.210492 & 0.4847392   &   0.4847392  & -3.349263  &
    -3.012227\\
    $\underline{\beta}_3$  & -6.010696 & -0.5739216   &  -0.5739216  &  5.531043  &
    -3.338545\\
    \bottomrule
    \multicolumn{6}{c}{\bf Std. Errors}\\
    \toprule
    \ & (Intercept) & Economics & Education & Health  & Political\\ \midrule
    $\underline{\beta}_1$  &  21.33983 & 1.569322   &    1.569322 &   21.81210  &
    2.775914\\
    $\underline{\beta}_3$  &  25.41046 & 1.976516   &   1.976516  &  25.93012   &
    3.562501\\
    \bottomrule
    \multicolumn{6}{c}{\bf Wald test pvalues}\\
    \toprule
    \ & (Intercept) & Economics & Education & Health  & Political\\ \midrule
    $\underline{\beta}_1$ &   0.9547646 & 0.7574095    &  0.7574095  &  0.8779640  &
    0.2778642\\
    $\underline{\beta}_3$ &   0.8130105 & 0.7715330    &  0.7715330  &  0.8310885  &
    0.3486891\\
    \bottomrule
    \end{tabular}
    \end{center}
    \caption{Estimates of the $\underline{\beta}_1, \underline{\beta}_3$ coefficients
    for {\bf row 2} of the transition matrix. The Wald test is testing the null hypothesis
    that the corresponding coefficient is equal to zero. Using a significance level
    of $90\%$, only the coefficients with a p-value below 0.1 are significant
    (i.e. the null hypothesis is rejected)}\label{row2}
\end{table}
Unfortunately, all the Wald test p-values here are bigger than 0.1, meaning that
we cannot find any significant relationship between the the proposed indicators and the probability for a country to change
its wage gap level starting in level 2.

\ 

\noindent{\bf Row 3 (Level 3 countries).}
The three elements $p_{31}(t),p_{32}(t),p_{33}(t)$ of the third row of the
transition matrix represent the probabilities that a country in level 3
(the worst level) at time $t$ passes to level 1,2,3, respectively,
at time $t+1$. Actually, as it was observed for row 1, looking at the estimated
overall transition matrix (\ref{overallP}) and also the single estimated
transition matrices, reported in the previous section, we can fix $p_{31}(t)=0$
for all considered years $t$. Thus, in the model (\ref{multimodel}) only the two
probabilities $p_{32}(t)$ and $p_{33}(t)$ survive, and, similarly to row 1, the
model reduces to the usual logistic regression model

$$
\ln \frac{p_{32}(t)}{p_{33}(t)}=\underline{\beta}\cdot \underline{Z}(t),
$$

where we have to estimate only one vector $\underline{\beta}$ of coefficients,
whose elements tell us the contribution of each covariate in the (relative)
probability of {\bf decreasing} the wage gap from level 3 to level 2. We
obtained the results reported in Table \ref{row3}.

\begin{table}[h]
    \begin{center}
    \begin{tabular}{lrrr}
    \toprule
    {\bf Coefficients} &  {\bf Values} &  {\bf Std. Err.} & {\bf Wald test pvalues}\\\midrule
    Intercept          &  30.60851927  & 21.868184        &  0.1616083\\
    Economics          &    1.78943095 & 1.117053         & 0.1091731 \\
    Education          &  1.78943095   & 1.117053         & 0.1091731 \\
    Health             &  -34.45431736 & 21.218702        & 0.1044246 \\
    Political          &   -0.03409214 & 2.505859         & 0.9891451 \\\bottomrule
    \end{tabular}
    \end{center}
    \caption{Estimatex of the $\underline{\beta}$ coefficient for {\bf row 3} of
    the transition matrix. The Wald test is testing the null hypothesis that the
    corresponding coefficient is equal to zero. Using a significance level of
    $90\%$, only the coefficients with a p-value below 0.1 are significant
    (i.e. the null hypothesis is rejected)}\label{row3}
\end{table}
We observe that the coefficients of the Economics, Education and Health
indicators are only slightly significant. Also, the coefficients of Economics and
Education are positive, while the coefficient of Health is negative. It, thus,
seems that for countries which start from level 3 (the worst
level), an improvement in the women/men employment rate and in the women/men
higher education is associated with an increase of the probability that the
country improves its situation, passing to level 2, while an
increase in the women/men life expectancy is reducing the probability that the
country passes to level 2.

Overall, we observe that the significance of our results is  low,
due to the lack of data. Our results and their interpretation could be improved in further work by collecting data from more countries and/or across more years.

\subsubsection{A latent Markov model for the wage gap indicator}
Wage gap indicator represents the quantification of an unobserved (latent) variable, which describes the disparities between women and man in the work environment. Being so, the wage gap of the $i$-th country at time $t$, $X_{it}$, is the observable outcome of the latent variable $U_{it}$.

As before, we assume that the wage gap of the $i$-th country is measured  over a sequence of $T$ years, $\underline{X}_{i}=(X_{i1}\ldots,X_{ik})$, where $i=1,\ldots,n$, $n=28$, and $k=8$ (years from 2009 until 2016). The main assumption of the basic Latent Markov Model (LMM) is  local independence \cite{Bartolucci.etal:2012}, i.e. for the $i$-th country, the response variables in $\underline{X}_{i}$ are conditionally independent given a latent process $\underline{U}_{i}=(U_{i1}\ldots,U_{ik})$. The latent process is assumed to follow a first-order Markov chain with state spaces ${1,\ldots,l}$ \cite{Bartolucci.etal:2010}. 

The model parameters are:
\begin{itemize}
	\item probability of each latent state, $\pi_u=P(U_{i1}=u)$, $u=1,\ldots,l$;
	\item transition probabilities between latent states, $\pi^{(t)}_{v|u}=P(U_{it}=v|U_{i,t-1}=u)$, $t=2,\ldots,k$, $u,v=1,\ldots,l$;
	\item conditional probability of wage gap be at  state $x$ given that the latent variable is at state $u$, $\phi^{(t)}_{x|u}=P(X_{it}=x|U_{it}=u)$, $t=1,\ldots,k$, $u=1,\ldots,l$, and $x=1,\ldots,m$.
\end{itemize} 
As before, we consider three stages for the wage gap, $m=3$, and the optimal choice of the number of latent stages is also three, $l=3$, according to the estimates obtained using the R package LMest \cite{LMest:2017}. 

As a result of adjusting the model to the dataset, the estimated initial  probabilities of each latent state are: 0.3668 for $u=1$, 0.2427 for $u=2$, and 0.3906 for $u=3$. The estimated conditional probabilities, summarized in Table \ref{CondRespProb}, allow us to say that for a country in $i$-th latent class, there is a high probability of the country belonging to the $i$-th wage gap class. Thus, the latent classes can be interpreted as ``Small'', ``Medium'', and ``High'' gender disparities in employment. Moreover, a country with High (Small) gender disparities in employment cannot have a Small (High) wage gap.

\begin{table}[h]
	\centering
	\small
	\caption{Estimate of the conditional probability of wage gap be at a certain state  given a latent state.}\label{CondRespProb}
	\begin{tabular}{lrrrr}
		\toprule
		 & & \multicolumn{3}{c}{Latent classes}\\\cline{3-5}
		 & & {\bf 1} &  {\bf 2} & {\bf 3}\\ \midrule
		     &Small & 0.985& 0.000& 0.000\\
   Wage Gap  &Medium& 0.015& 0.982& 0.000\\
		     &High  & 0.000& 0.0128& 1.000\\	
	\bottomrule
	\end{tabular}

\end{table}

The obtained estimated transition probability matrix is:
\begin{equation}
P_{LMM}=\begin{bmatrix}
     0.9369& 0.0631& 0.0000\\
     0.0614& 0.9040& 0.0346\\
     0.0000& 0.1122& 0.8878\\
\end{bmatrix}. \label{TransP_LMM}
\end{equation}
The estimated transition matrix leads to similar results and conclusions reached in Section \ref{Sec:MM}, when a Markov model is adjusted to the data. Thus, we will not repeat the model fitting discussed in Section \ref{Sec:Log_Reg}.  In Table \ref{Patterns}, we list the over time state patterns associated with each country as well as the most frequent state during the period under study. The latent class with highest a posteriori probability coincide with the observed class at a certain year. As a consequence, countries can be grouped by the mode state, which is the order appearing in Table \ref{Patterns}.

\begin{table}[h]
	\centering
	\small
	\caption{Wage gap patterns over the years and the associated most frequent state (mode).}\label{Patterns}
	\begin{tabular}{lrrrrrrrr|c}
		\toprule
		 & \multicolumn{8}{c|}{Wage Gap classes}&Mode\\\toprule{2-9}
  Country    &2009& 2010& 2011& 2012& 2013& 2014& 2015& 2016&state\\\midrule
\begin{tabular}{l}
Belgium, Italy \\ Luxembourg, Malta \\ Poland, Romania\\ Slovenia\\
\end{tabular}            
              &1&    1&    1&    1&    1&    1&    1&    1&1\\
Lithuania      &2&    1&    1&    1&    1&    1&    2&    2&1\\
Portugal      &1&    1&    1&    2&    1&    2&    3&    2&1\\


\begin{tabular}{l}
	Denmark, France\\Norway\\ 
\end{tabular}
               &2&    2&    2&    2&    2&    2&    2&    2&2\\
Bulgaria       &1&    1&    1&    2&    2&    2&    2&    2&2\\
Latvia         &1&    2&    2&    2&    2&    2&    2&    2&2\\
Spain          &2&    2&    2&    3&    3&    2&    2&    2&2\\
Sweden         &2&    2&    2&    2&    2&    1&    1&    1&2\\
Cyprus         &3&    2&    2&    2&    2&    2&    1&    1&2\\
Netherlands    &3&    3&    3&    2&    2&    2&    2&    2&2\\
Switzerland    &3&    3&    2&    2&    2&    2&    2&    2&2\\
Hungary        &2&    2&    3&    3&    3&    2&    1&    1&2, 3\\
\begin{tabular}{l}
	Czechia,   Germany\\       
	Estonia,   Austria\\
	Slovakia, UK\\ 
\end{tabular}
               &3&    3&    3&    3&    3&    3&    3&    3&3\\
Finland        &3&    3&    3&    3&    3&    3&    2&    2&3\\
Iceland        &3&    3&    3&    3&    3&    2&    2&    2&3\\
\bottomrule
	\end{tabular}
	
\end{table}

%
%

Given the proximity between the estimated transition matrices \eqref{overallP} and \eqref{TransP_LMM}  and the identity matrix, we suspect that wage gap time dependency is not well captured by the data. This fact is perfectly justified by the small number of time periods under study. Partial autocorrelation functions evaluated per country confirmed this claim. In the next section, we analyse the data ignoring potential time dependencies.

\subsubsection{Ranking countries based on gender disparities}

\begin{figure}
	\centering
	\includegraphics[width=.5\textwidth]{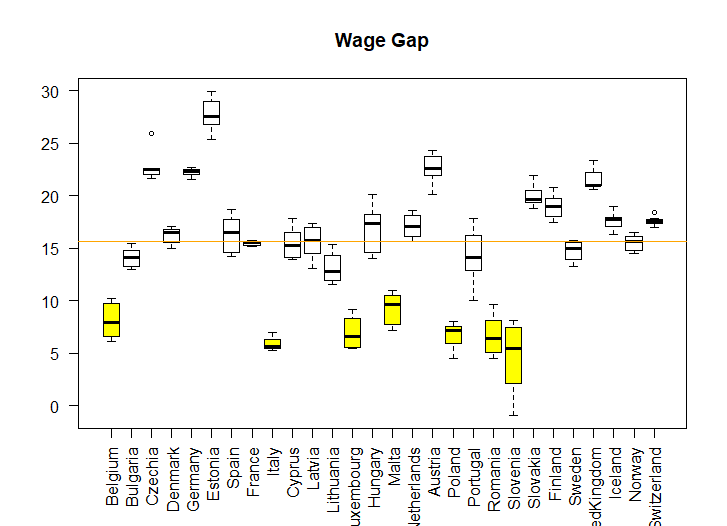}
	\caption{Wage Gap, in yellow are the countries with the lowest wage gap. The orange line marks the overall wage gap median.}\label{fig:Original_WG}
\end{figure}

\begin{figure}
	\centering
	\begin{flushleft}
		\captionsetup{justification=raggedright}
		\hspace*{5pt}
		{\label{fig:Ex1_Conv}\includegraphics[width=.45\textwidth]{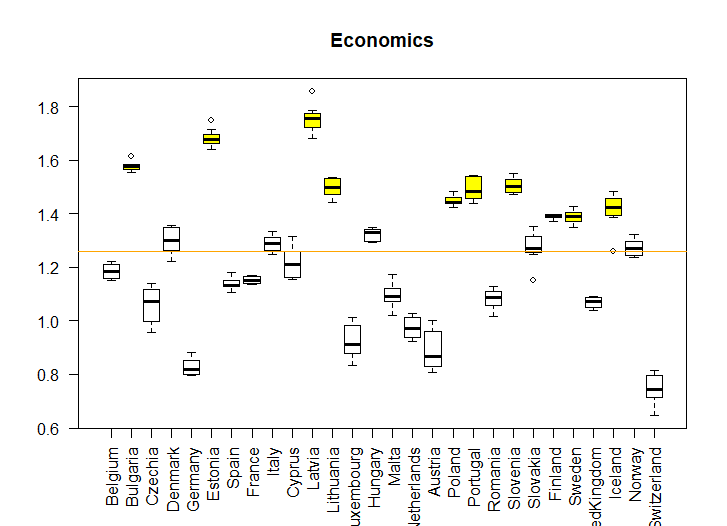}}
		{\label{fig:Ex1_Syn}\includegraphics[width=.45\textwidth]{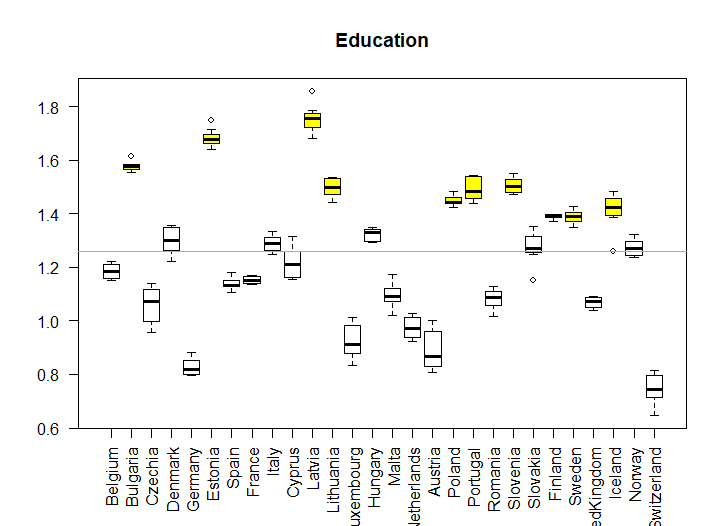}} 
	\end{flushleft}              
	\caption{Economic and Education, in yellow are the countries with the highest value on economical and educational index, respectively. The orange line marks the overall median.}\label{fig:Original_Econ&Educ}
\end{figure}

\begin{figure}
	\centering
	\begin{flushleft}
		\captionsetup{justification=raggedright}
		\hspace*{5pt}
		{\label{fig:Ex1_Conv}\includegraphics[width=.45\textwidth]{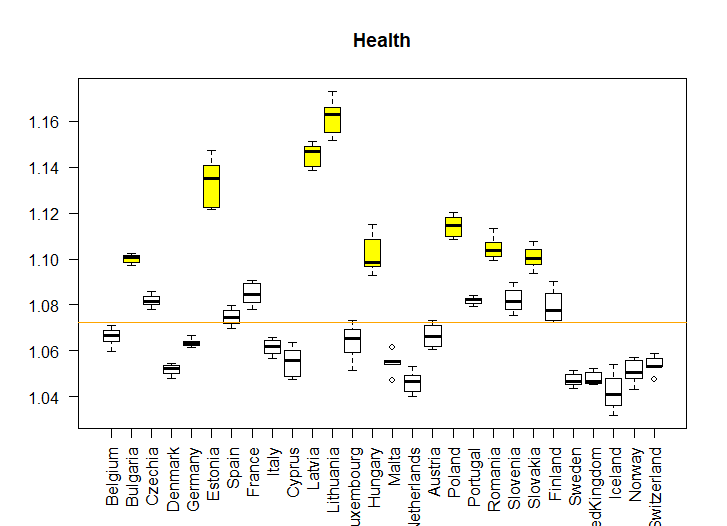}}
		{\label{fig:Ex1_Syn}\includegraphics[width=.45\textwidth]{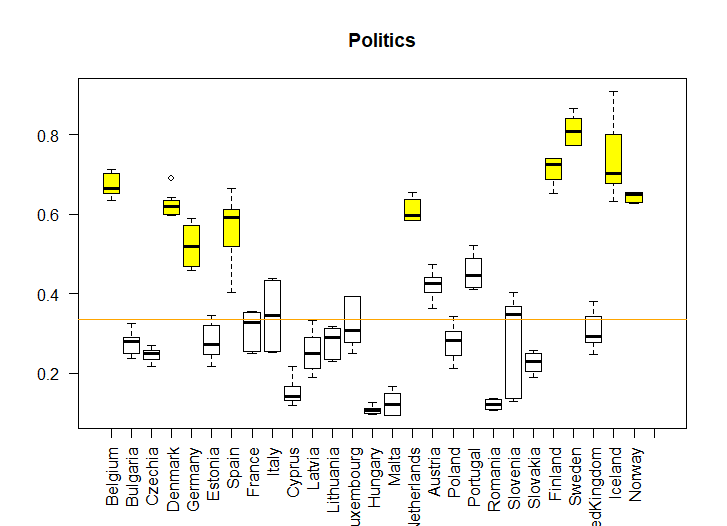}} 
	\end{flushleft}              
	\caption{Health and Politics, in yellow are the countries with the highest value on health and on women political participation index, respectively. The orange line marks the overall median.}\label{fig:Original_health&Pol}
\end{figure}

Gender gap can be quantified in many ways and several features can be used to construct such indexes. The variables under study are described in Section \ref{Sec:ModelMC}, and the boxplots of their values over the years under study are shown in Figures \ref{fig:Original_WG} to \ref{fig:Original_health&Pol}. Countries with the lowest gender gap or most favorable female-to-male ratios to women are marked in yellow and the orange line represents the overall median. According to Figure \ref{fig:Original_WG},  countries with the lowest wage gap are Belgium, Italy, Luxembourg, Malta, Poland Romania, and Slovenia. In fact, Slovenia registers a negative value in 2009, meaning that women over that year had a higher mean daily salary than men. 
Countries with the highest wage gap are Estonia, followed by Czechia, Germany, and Austria.

The political participation only registers values lower than one so all European parliaments over the years had more men than women. The countries with the lowest gap in terms of women and man political participation are Belgium, Finland, Sweden, Iceland, and Norway. The second group of countries with a median female-to-male ratio over 0.5 (0.5 means one woman per two men in the parliament) is formed by Denmark, Germany, Spain, and the Netherlands. The countries that register the highest discrepancy between women and men political participation are Cyprus, Hungary, Malta, and Romania.

The Economics and Education index show very similar patterns. In fact, these two variables are highly correlated (higher than 0.969) and from now on the only economic variable is going to be considered. The economic index measures the female-to-male ratio of employment rates of highly educated individuals. Thus, the generality of the countries have rates higher than 1. The exceptions are Czechia, Germany, Luxembourg, Netherlands, Austria, and Switzerland. The countries that register the highest values are Bulgaria, Estonia, and Latvia followed by Lithuania, Poland, Portugal, Slovenia, Finland, and Iceland (vide Figure \ref{fig:Original_Econ&Educ}).

The health index is measured by the female-to-male ratio of life expectancy. According to Figure \ref{fig:Original_health&Pol}, all countries register values higher than one, meaning that women life expectancy is higher than men. Countries that have the largest differences between women and men life expectancy are Estonia, Latvia, and Lithuania followed by Bulgaria, Hungary, Poland, Romania, and Slovakia. This index clearly raises the question about parity between gender. Countries like Netherlands, Sweden, United Kingdom, Iceland or Norway register ratios near 1. And because of that, they are the ones where the life expectancy of the two genders are more alike.

In the Gender Gap Report 2017 \cite{WEF}, produced by the World Economic Forum, the authors argue that female-to-male ratios measure women's empowerment. Equality between genders is obtained when the ratio is equal to one, so they truncated the ratios at a given ``equality benchmark''. This strategy assigns the same scores to a country that has reach parity between women and men or to a country where women have surpassed men \cite{WEF}.  Thus, if $X_i$ represents the wage gap as defined in Section \ref{Sec:MM}, the new wage index is just a ratio of female-to-male daily mean salary $Y_{1i}=(1-X_i/100)$. Note that there is not a need to truncate this variable since for all the countries under study the observed values are always smaller than one, meaning that none of the countries reached parity (with the exception of Slovakia in 2009).  
Similarly, political participation also registers ratios lower than one. The equality benchmark   for life expectancy is $1.06\simeq 87.5/82.5$, following the Gender Gap Report 2017 \cite{WEF} recommendation. For comparison reasons, this index is also transformed in values between 0 and 1. Being so, if $Z_2$ represents the health index, the new variable is $Y_{2}=\min{\left(Z_2,1.06\right)}/1.06$. The equality benchmark value for Economics and Education is set to one.

\begin{figure}
	\centering
	\begin{flushleft}
		\captionsetup{justification=raggedright}
		\hspace*{5pt}
		{\label{fig:Ex1_Conv}\includegraphics[width=.45\textwidth]{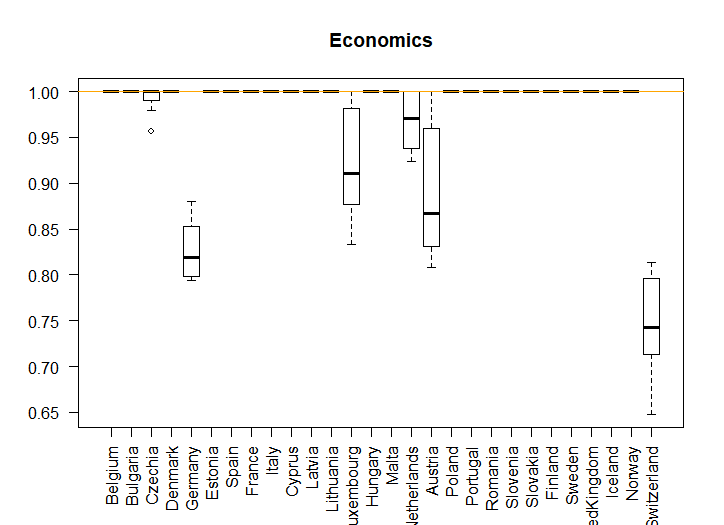}}
		{\label{fig:Ex1_Syn}\includegraphics[width=.45\textwidth]{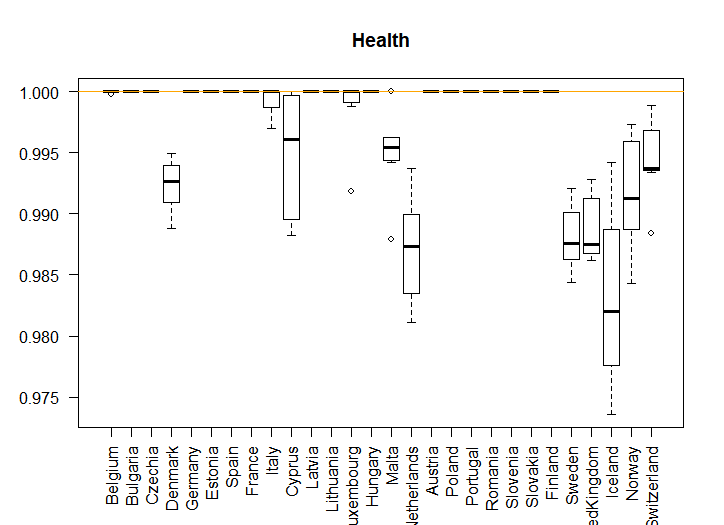}} 
	\end{flushleft}              
	\caption{Truncated indexes of Economic and Health. Value one represents parity between genders.}\label{fig:Trunc_Econ&Health}
\end{figure}

In Figure \ref{fig:Trunc_Econ&Health}, the truncated indexes for Economic and Health are shown. For these indexes,  1.0  represents parity and low values (approaching zero) represent high disparities between genders. Figure \ref{fig:Trunc_Econ&Health} highlights the parity scenarios from European countries related to the chosen economic and health indexes. Major disparities between genders in terms of employment rates of highly educated individuals are observed in countries like Germany, Luxembourg, Netherlands, Austria, and Switzerland. From the health point of view, Iceland has the highest variability along the years and register the lowest values on this index.

To rank the countries we used two different sets of variables. In the first dataset, we only considered the truncated wage gap values over the years: $(Y_{1,1},\ldots,Y_{1,k})$, where $Y_{1,j}$ represents the truncated wage gap in the $j$-th year and as before $k=8$. The second dataset was build by averaging the truncated wage gap, $Y_{1,j}$, economics, $Y_{2,j}$, health $Y_{4,j}$, and political, $Y_{5,j}$, i.e. we analyse $(W_{1},\ldots,W_{k})$, where $W_{j}=(Y_{1,j}+Y_{2,j}+Y_{4,j}+Y_{5,j})/4$. To each of these datasets, we estimate the classical principal components. In both cases, the first principal component explains the vast majority of the observed variability (95.4\% for the first dataset and  95.0\% for the second one). The first principal components, based on each dataset, are defined by:

\begin{eqnarray*}
	PC_{1,1}=&0.375Y_{1,1}+ 0.375Y_{1,2}+ 0.357Y_{1,3}+ 0.370Y_{1,4}+\\
	         &0.358Y_{1,5}+ 0.346Y_{1,6}+ 0.328Y_{1,7}+ 0.314Y_{1,8},&\quad {\rm for\, dataset\, 1,}\\
	PC_{1,2}=&0.371W_{1,1}+ 0.353W_{1,2}+ 0.349W_{1,3}+ 0.353W_{1,4}+\\
	         &0.338W_{1,5}+ 0.348W_{1,6}+ 0.353W_{1,7}+ 0.362W_{1,8}+,&\quad {\rm for\, dataset\, 2.}
\end{eqnarray*}

\begin{figure}
	\centering
	\includegraphics[width=.7\textwidth]{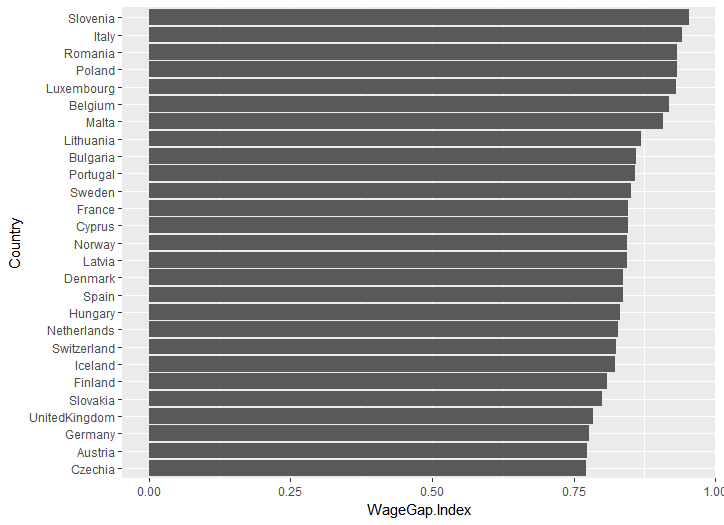}
	\caption{Overall wage parity index by country. The value 1.0 represents parity between genders.}\label{fig:Rank_WG}
\end{figure}

\begin{figure}
	\centering
	\includegraphics[width=.7\textwidth]{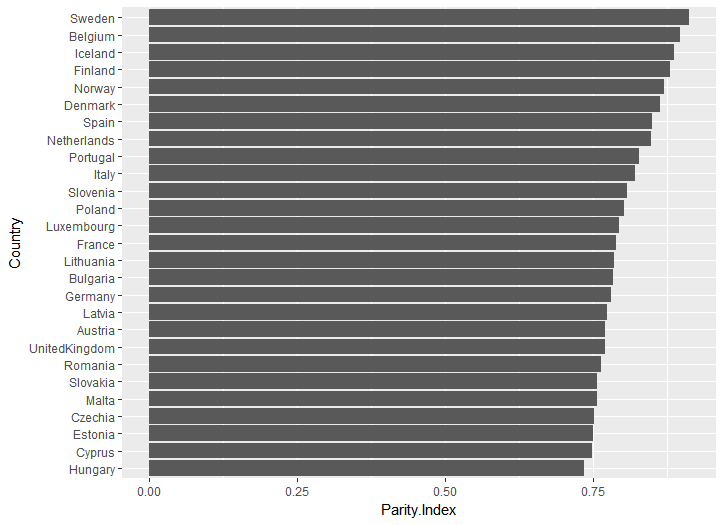}
	\caption{Overall gender parity index by country.  The value 1.0 represents parity between genders.}\label{fig:Rank_All}
\end{figure}

\begin{figure}
	\centering
	\begin{flushleft}
		\captionsetup{justification=raggedright}
		\hspace*{5pt}
		\subfloat[Wage gender gap.]
		{\label{fig:Ex1_Conv}\includegraphics[width=.48\textwidth]{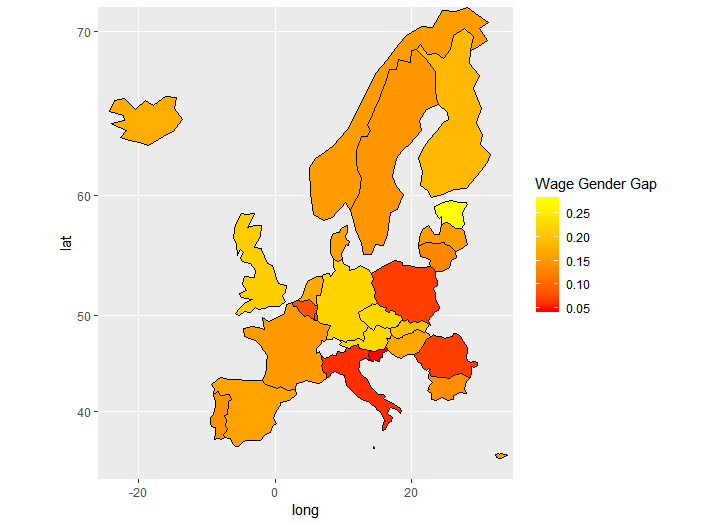}}\label{Map_WageGenderGap}
		\subfloat[Overall Gender gap.]
		{\label{fig:Ex1_Syn}\includegraphics[width=.48\textwidth]{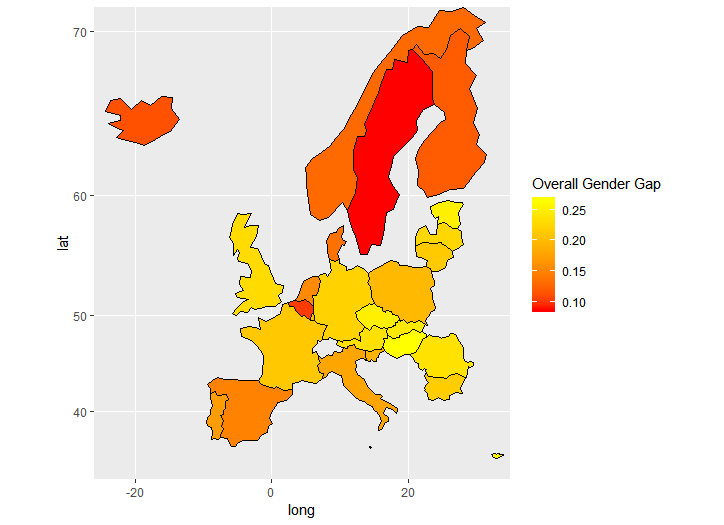}}\label{Map_WageGenderGap} 
	\end{flushleft}              
	\caption{Gender gap indexes, on the left based on wage female-to-male ratios and on the right based on all the dimensions under study. The value 0 represents parity between genders.}\label{fig:Map_GenderGap}
\end{figure}

These principal components can be interpreted as a weighted average of the indexes observed from 2009 to 2016. Thus, high values on the first principal components mean countries with lower wage gap (or gender gap, accordingly with the dataset under consideration) and describe a situation nearer parity between genders. 
Nevertheless, the maximum possible value of these indexes is not one. A country with parity in all the years under study would have all the indexes equal to one, so to transform the scores into values between 0 and 1 we just have to divide the observed scores by the ``parity score'', 2.827 for dataset 1 and 2.824 for dataset 2, values obtained by considering $Y_{1,j}=1$ and $W_{j}=1$, for $j=1,\ldots,8$, respectively. The sorted ranks according to the two approaches are illustrated in Figure \ref{fig:Rank_WG} and \ref{fig:Rank_All}. If we consider all the features to rank the countries, Belgium and the Nordic countries are the ones with lower gender gap (vide Figure \ref{fig:Rank_All}). Nevertheless, if we only consider wage female-to-male ratios over the 8 years under study (vide Figure \ref{fig:Rank_WG}), the best six positions are occupied by Slovenia, Italy, Romania, Poland, Luxembourg, and Belgium. The only country that stands by having a low gender gap, independently from the variables considered, is Belgium.  
Hungary, Cyprus, and Estonia have the lowest parity index when all the variables are taken into consideration, while the worst three countries are Czechia, Austria, and Germany when only wage ratios are used.  
For a better comparison between ranks, Figure \ref{fig:Map_GenderGap} represents the Gender gap index, defined as one minus the corresponding parity index.

In the next subsection variables  $(Y_{1,1},\ldots,Y_{1,8})$ and $(W_{1},\ldots,W_{8})$ are used to build clusters of homogeneous countries.

\subsubsection{Clustering countries based on gender disparities}

In order to obtain groups of countries with similar gender parity based on the two different datasets, we tested several clustering methods and dissimilarities. We concluded to using the partition around medoids method \cite{Kaufman.Rousseeuw:1990} which considers the Euclidean distance as the dissimilarity measure between countries. We elect three clusters given that the discretization of the wage gap in Section \ref{Sec:ModelMC}  also considered three classes. 

\begin{figure}
	\centering
	\begin{flushleft}
		\captionsetup{justification=raggedright}
		\hspace*{5pt}
		\subfloat[Wage gender gap.]
		{\includegraphics[width=.48\textwidth]{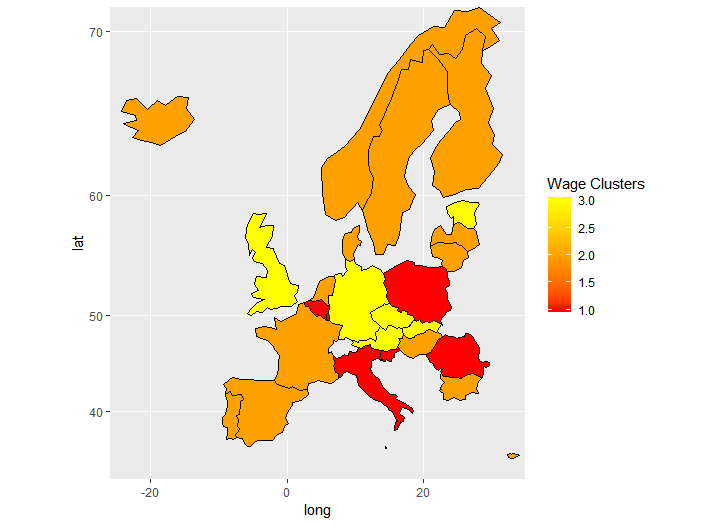}}\label{Map_WageClusters}
		\subfloat[Overall Gender gap.]
		{\includegraphics[width=.48\textwidth]{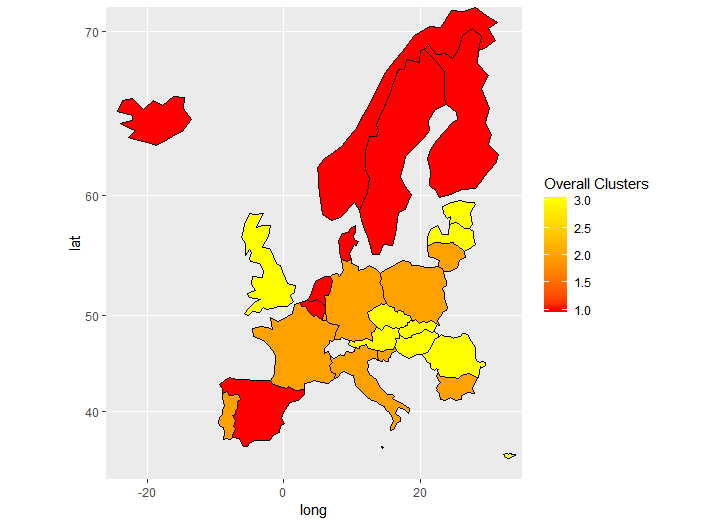}}\label{Map_OverallClusters} 
	\end{flushleft}              
	\caption{Clustering partitions, on the left based on wage female-to-male ratios and on the right based on all the dimensions under study. Cluster 1 (red) contains countries with low gender gap, cluster 2 (orange) is formed by countries with medium gender gap, and cluster 3 (yellow) contains the countries with high gender gap.}\label{fig:Map_Clusters}
\end{figure}

The interpretation of the two partitions follows similar reasoning: countries are grouped by the gender gap level. Thus, cluster 1 contains the countries with the narrower gender gap, cluster 2 is formed by countries with a medium gender gap level, and finally, cluster 3 is formed by countries with the highest gender gap. The list of the two partitions is available in Table \ref{Tab:Clusters}. Even though, the partitions are based on different characteristics, the cases that are assigned to the same level of gender gap (marked in bold in Table \ref{Tab:Clusters}) are: Belgium (lower gender gap), Bulgaria, France, Lithuania, and Portugal (Medium level) and Austria, Czechia, and Estonia (high gender gap). Another important remark is that the Nordic countries are all assigned as having low gender gap levels when all the dimensions are considered but are not so well classified when only wage indicators are taken into consideration. 
These considerations turn to be clearer, when we associate a colour to each cluster and visualize countries in the European map accordingly, as shown in Figure \ref{fig:Map_Clusters}.

\begin{table}[h]
	\centering
	\small
	\caption{Gender gap cluster of countries based on the wage female-to-male ratios over the years (Wage-specific) and obtained when the wage, health, economic, and politic indicators are all taken into consideration (Overall).}\label{Tab:Clusters}
	\begin{tabular}{l|rrr}
		\toprule
		Country	&  \multicolumn{2}{|c}{Gender Gap Level}\\\toprule{2-3} &
		{Wage-specific} &  {Overall} \\ \midrule \textbf{Belgium}	&	Low	&	Low	\\
Denmark	&	Medium	&	Low	\\
Finland	&	Medium	&	Low	\\
Iceland	&	Medium	&	Low	\\
Netherlands	&	Medium	&	Low	\\
Norway	&	Medium	&	Low	\\
Spain	&	Medium	&	Low	\\
Sweden	&	Medium	&	Low	\\
Italy	&	Low	&	Medium	\\
Luxembourg	&	Low	&	Medium	\\
Poland	&	Low	&	Medium	\\
Slovenia	&	Low	&	Medium	\\
\textbf{Bulgaria}	&	Medium	&	Medium	\\
\textbf{France}	&	Medium	&	Medium	\\
\textbf{Lithuania}	&	Medium	&	Medium	\\
\textbf{Portugal}	&	Medium	&	Medium	\\
Germany	&	High	&	Medium	\\
Malta	&	Low	&	High	\\
Romania	&	Low	&	High	\\
Cyprus	&	Medium	&	High	\\
Hungary	&	Medium	&	High	\\
Latvia	&	Medium	&	High	\\
Slovakia	&	Medium	&	High	\\
UnitedKingdom	&	Medium	&	High	\\
\textbf{Austria}	&	High	&	High	\\
\textbf{Czechia}	&	High	&	High	\\
\textbf{Estonia}	&	High	&	High	\\
Switzerland	&	Medium	&		\\
		\bottomrule
	\end{tabular}
\end{table}

We cross the parity indexes with the two partitions in order to have a more clear interpretation and understanding of the obtained clusters. The first conclusion is that given the very low variability of the parity indexes describing the health and economic dimensions of the countries it does not seem they have differences among clusters. 

\begin{figure}
	\centering
	\captionsetup{justification=raggedright}
	\subfloat[Wage, Wage partition.]{\label{fig:MedianWG_ClusterWG}\includegraphics[width=.3\textwidth]{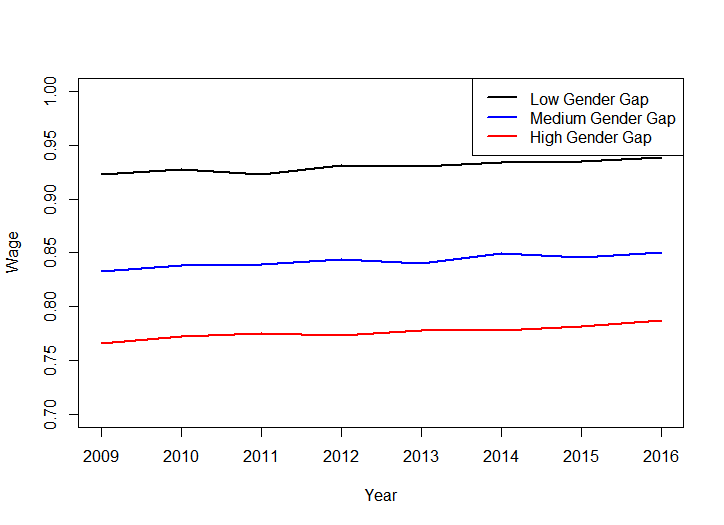}}              
	\hspace*{5pt}
	\subfloat[Politics, Wage partition.]{\label{fig:MedianPolitic_ClusterWG}\includegraphics[width=.3\textwidth]{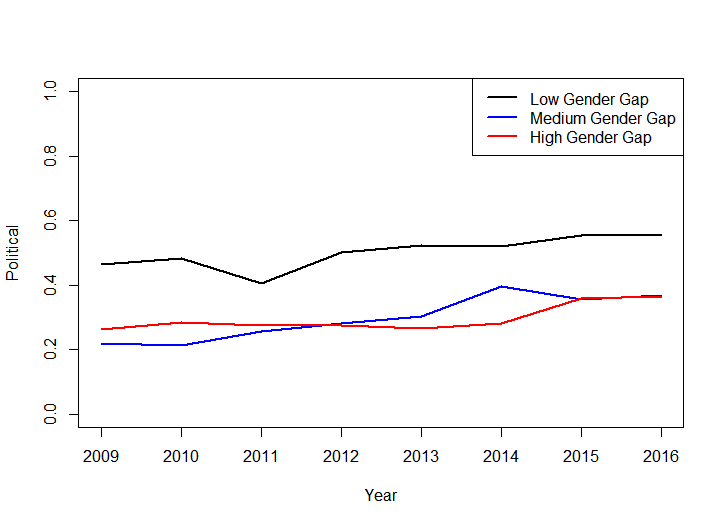}}
	\hspace*{5pt}
	\subfloat[Overall, Wage partition.]{\label{fig:MedianAll_ClusterWG}\includegraphics[width=.3\textwidth]{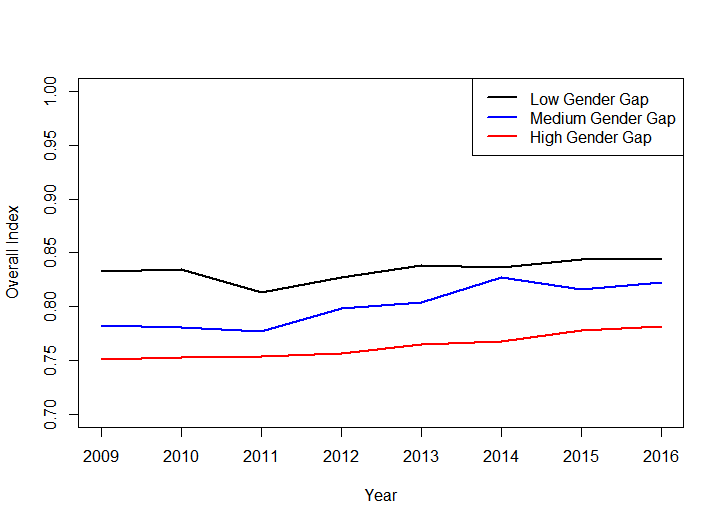}}\\
	\subfloat[Wage, Overall partition.]{\label{fig:MedianWg_ClusterOverall}\includegraphics[width=.3\textwidth]{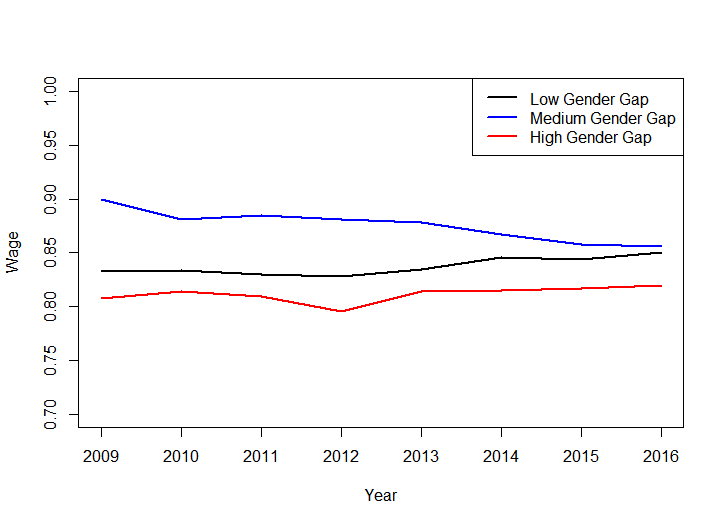}}
	\hspace*{5pt}
	\subfloat[Politics, Overall partition.]{\label{fig:MedianPolit_ClusterOverall}\includegraphics[width=.3\textwidth]{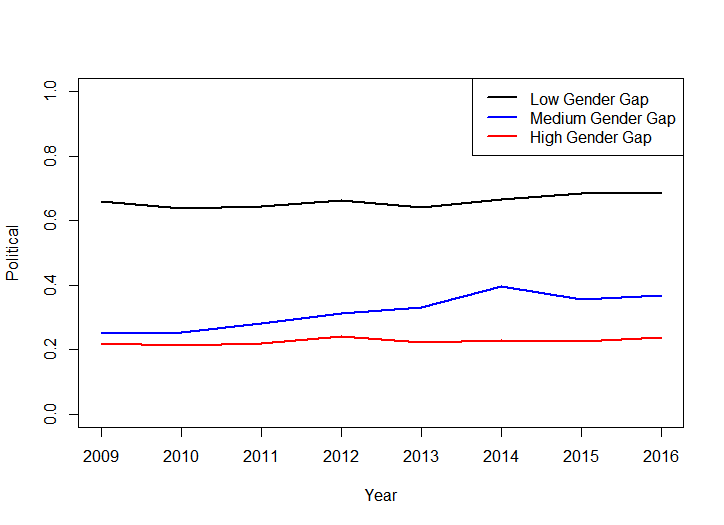}}
	\hspace*{5pt}
	\subfloat[Overall, Overall partition.]{\label{fig:MedianAll_ClusterOverall}\includegraphics[width=.3\textwidth]{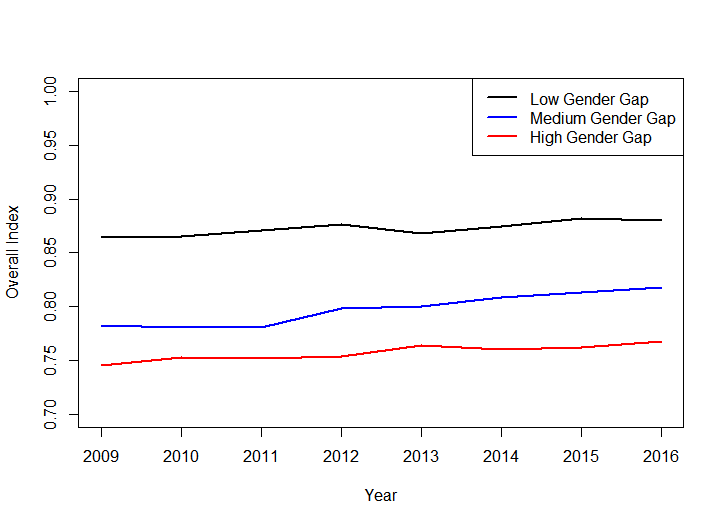}}          
	\caption{Yearly medians per cluster, first (second) row represents the partition based on the Wage (Overall) parity indexes.}
	\label{fig:MedianPartitions}
\end{figure}

In Figure \ref{fig:MedianPartitions} we show the yearly median cluster values of the parities indexes based on the two partitions. Figure  \subref*{fig:MedianWG_ClusterWG} and \subref*{fig:MedianAll_ClusterOverall} just confirm the clusters interpretation. When the partition was obtained based on the wage parity indexes (first row of Figure \ref{fig:MedianPartitions}), the clusters with medium and high wage gap levels show very similar (and low) political participation of women (vide Figure \subref*{fig:MedianPolitic_ClusterWG}). 

When the partition was estimated considering all the dimensions under study, the cluster with higher wage parity between genders is the one having medium overall gender gap (vide Figure  \subref*{fig:MedianWg_ClusterOverall}). It is a higher parity between gender on political participation that stands to characterize the low overall gender gap cluster, as Figure \subref*{fig:MedianPolit_ClusterOverall} illustrates. 

\captionsetup[subfigure]{subrefformat=simple,labelformat=simple,listofformat=subsimple}
\renewcommand\thesubfigure{(\arabic{subfigure})}

\newpage

\subsection{Analysing the academia in Cyprus: Women in STEM}
\label{section:cyprus_data}
\subsubsection{Introduction}
Over the last 60 years Cyprus has been through the difficult conditions of colonialism, independence, invasion and semi-occupation of its territory \cite{nearchou}. However, it managed to overcome these situations and to build up a strong economy and a competitive education system. Now it belongs to the European countries that has been developing rapidly during the last two decades from a poor country to a  high-income economy \cite{econ}. Building also its education system, Cyprus has the highest percentage of citizens of working age who have higher-level education in the EU at 30\%, and 47\% of its population aged 25–34 have tertiary education [REF wiki].  Despite this fast progress, the gender wage gap continues to be an important issue. During 2017, Cyprus ranked 20th in the EU on the Gender Equality Index with 56.3 out of 100 points and a score 11.1 points lower than the average EU score \cite{gendind}. Despite its faster improvement in the domains of health and survival, and economic participation and opportunity compared to other Member States of the EU, gender inequalities in education continue to persist. The gender inequality in education is largest in the STEM subjects, contributing significantly to the wage gap between men and women, as STEM jobs are on average highly paid and STEM skills are universally appealing and recognised. Hence, the focus of this section is to study the gender equality in the STEM fields in Cyprus which is a small, but yet critical region of EU, not only due to its geographical location, but also due to its sociopolitical and economical situation. We took a first step towards this direction in the Study Group week by analysing data provided by (i) the University of Cyprus (UCY), for the STEM fields, and (ii) the Statistical Service of Cyprus, for the number of Cypriots studying abroad and the population aged between 20-29.\\

\subsubsection{Preliminary analysis}

The University of Cyprus is a public research university established in 1989. It admitted its first students in 1992, and has now
approximately 7000 students.  It has become internationally recognized as a leading research institution
for its contribution to the advancement of science and culture, attracting external research funding and a number of
prestigious chairs of excellence. It is a member of several university Networks and
associations worldwide, ranking first among the top 10 national beneficiaries of
the EC financial contribution (according to the e-CORDA data update in July
2014) and has implemented successfully more than 95 EU projects during the past three
years. UCY is currently the largest research-intensive university in the country. 

Here, we aim to study gender equality in education and academic careers, in the STEM subjects. STEM subjects have been traditionally male-dominated, with
a historically low percentage of women since the Age of Enlightenment. 

Data
on the number of male and female students and academic staff members (permanent and non-permanent) were provided by the university. In the eight faculties of the University, we focused on the STEM subjects: (i) Pure and Applied Sciences, consisting of the departments of Biological Sciences, Mathematics and Statistics, Computer Science, Physics and Chemistry; and (ii) Engineering, consisting of the departments of Electrical and Computer Engineering, Mechanical and Manufacturing Engineering and Civil and  Environmental Engineering\footnote{The data were provided by the UCY prior to the Study Group, in
two files, one for academic staff and one for students, and were translated by the team
from Greek to English.} From these data the percentage of women in STEM subjects
by department (undergraduate, postgraduate) and the percentage of female faculty
members in all academic levels (Lecturer, Assistant Professor, Associate
Professor, Professor, and also non-permanent postdoctoral
researchers) in STEM subjects were analysed.  Through this preliminary analysis, we extracted some key trends and patterns and  compared the results with the EU average. Furthermore, to cover all aspects of gender inequality in Cyprus we analysed also data provided by the Statistical Service of Cyprus, about the students abroad and the population aged between 20-29.

\subsubsection{Analysis of STEM student data}
\begin{figure*}[!htbp] 
\subfloat[\label{grad_depart_a}]{%
       \includegraphics[width=.48\textwidth]{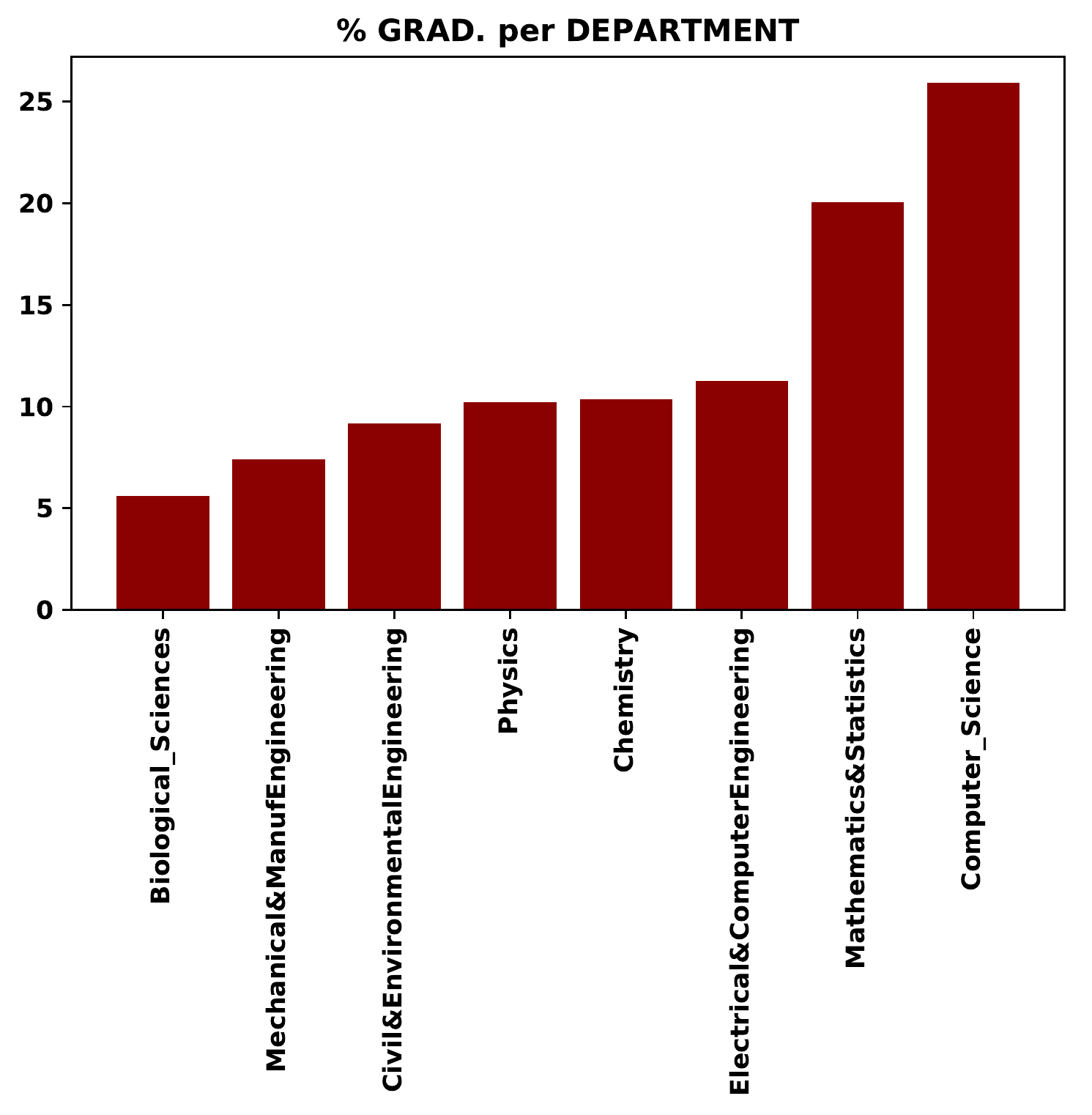}
     }
     \hfill
     \subfloat[\label{grad_depart_b}]{%
       \includegraphics[width=.48\textwidth]{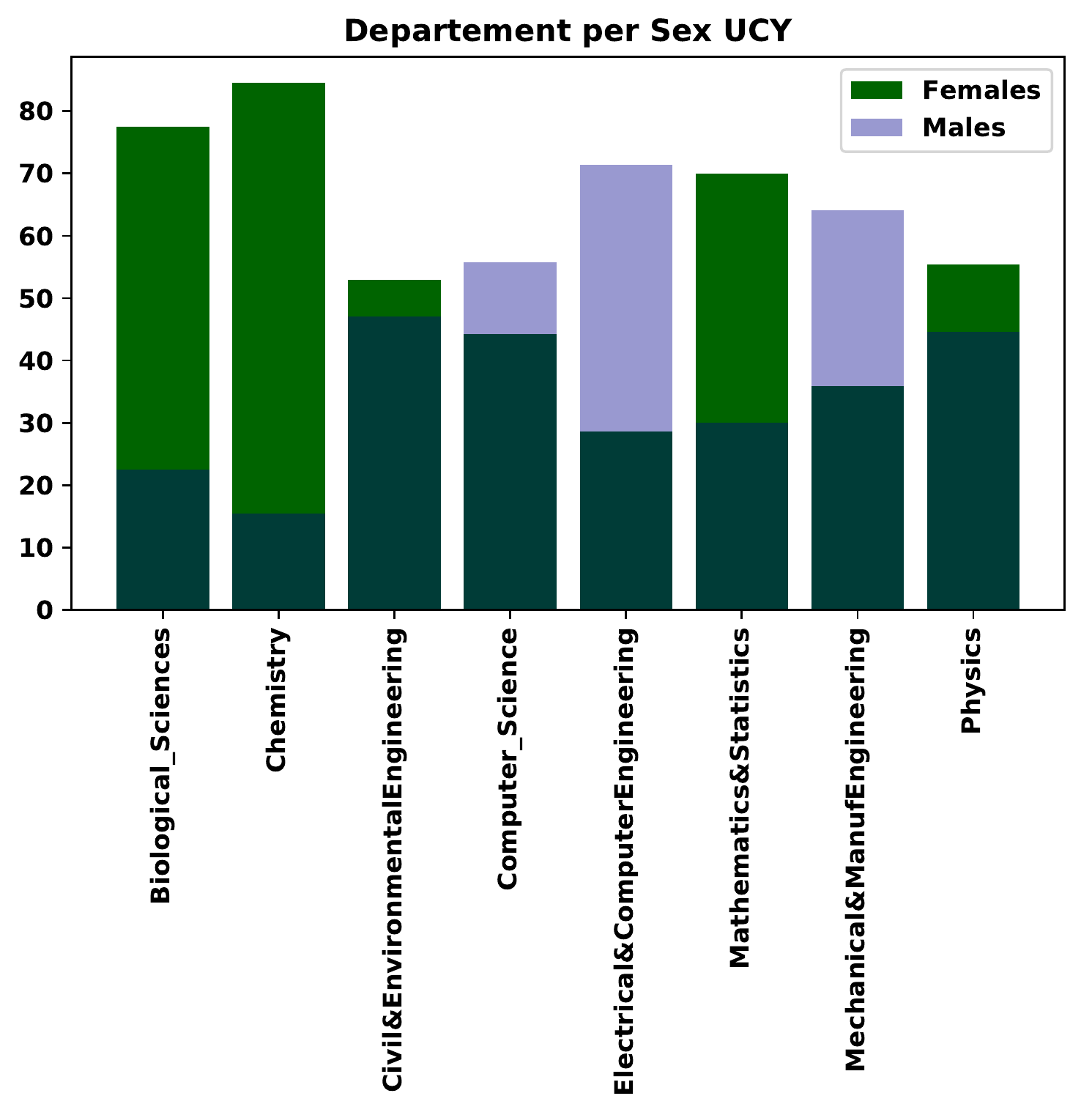}
     }
\caption{Students that completed a degree in a STEM subject at 
UCY, from 1996-2018, per department. (1) Percentage of male and female students. (2) Gender ratio of students.}
\label{grad_depart}
\end{figure*}

\begin{figure}[h!] 
    \centering
    \includegraphics[width=.8\textwidth]{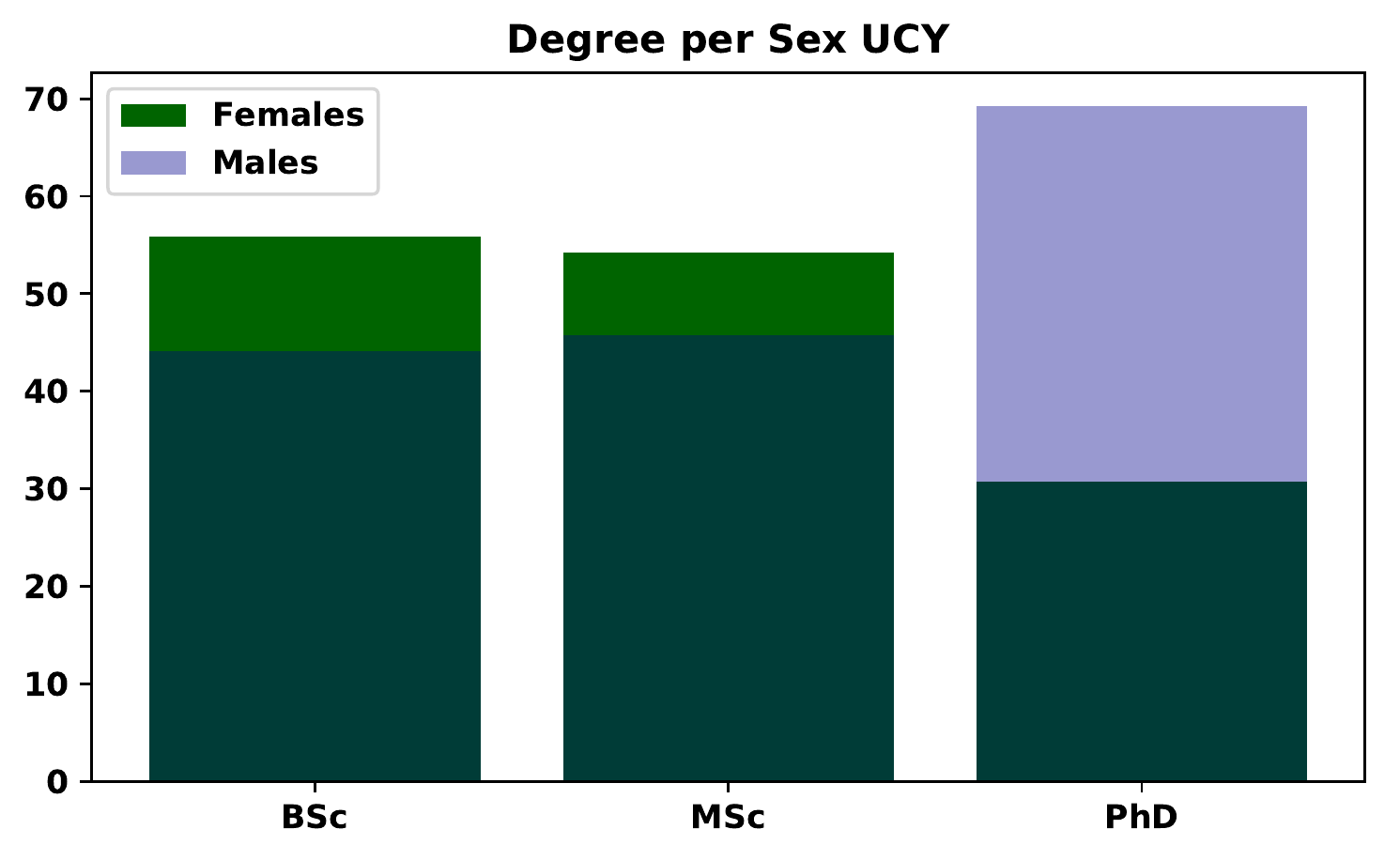}
    \caption{Gender ratio of students that completed a degree in a STEM subject at UCY, per degree.}
 \label{degree_per_sex}
\end{figure}

\begin{figure*}[!htbp] 
\centering
\subfloat[\label{Gender_eq_UoC_years_a}]{%
       \includegraphics[width=.48\textwidth]{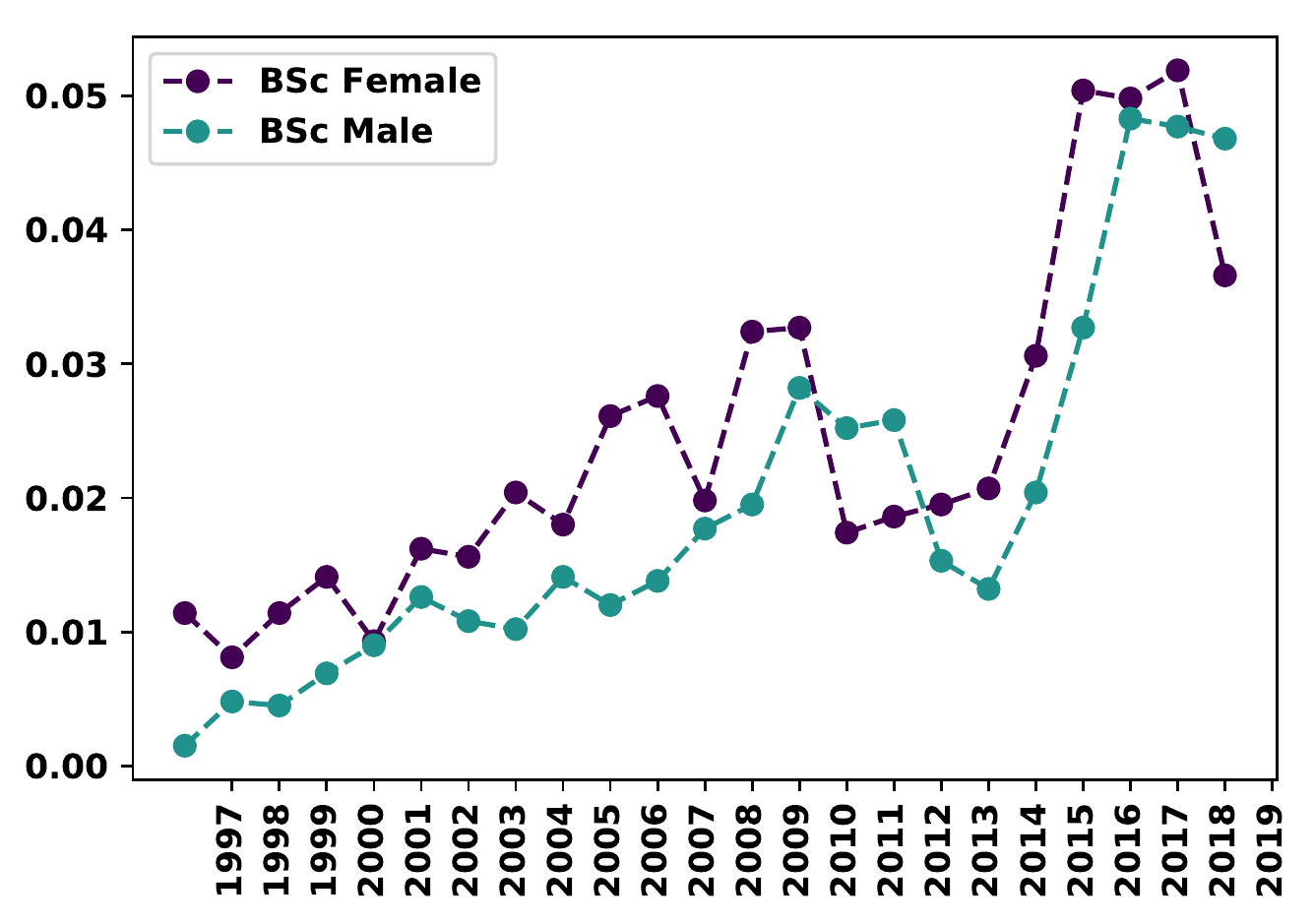}
     }
     \hfill
     \subfloat[\label{Gender_eq_UoC_years_b}]{%
       \includegraphics[width=.48\textwidth]{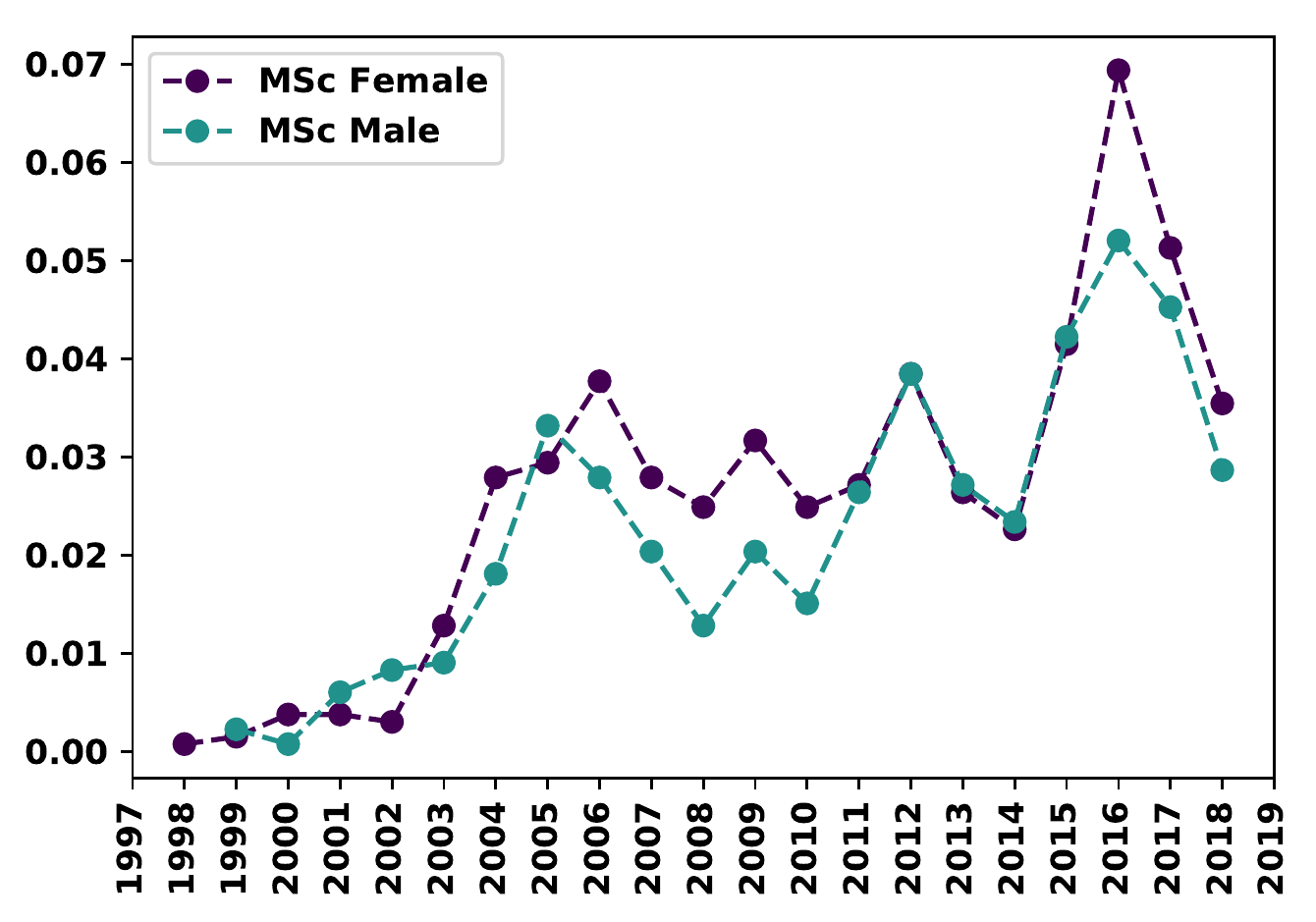}
     }\\
     \subfloat[\label{Gender_eq_UoC_years_c}]{%
       \includegraphics[width=.48\textwidth]{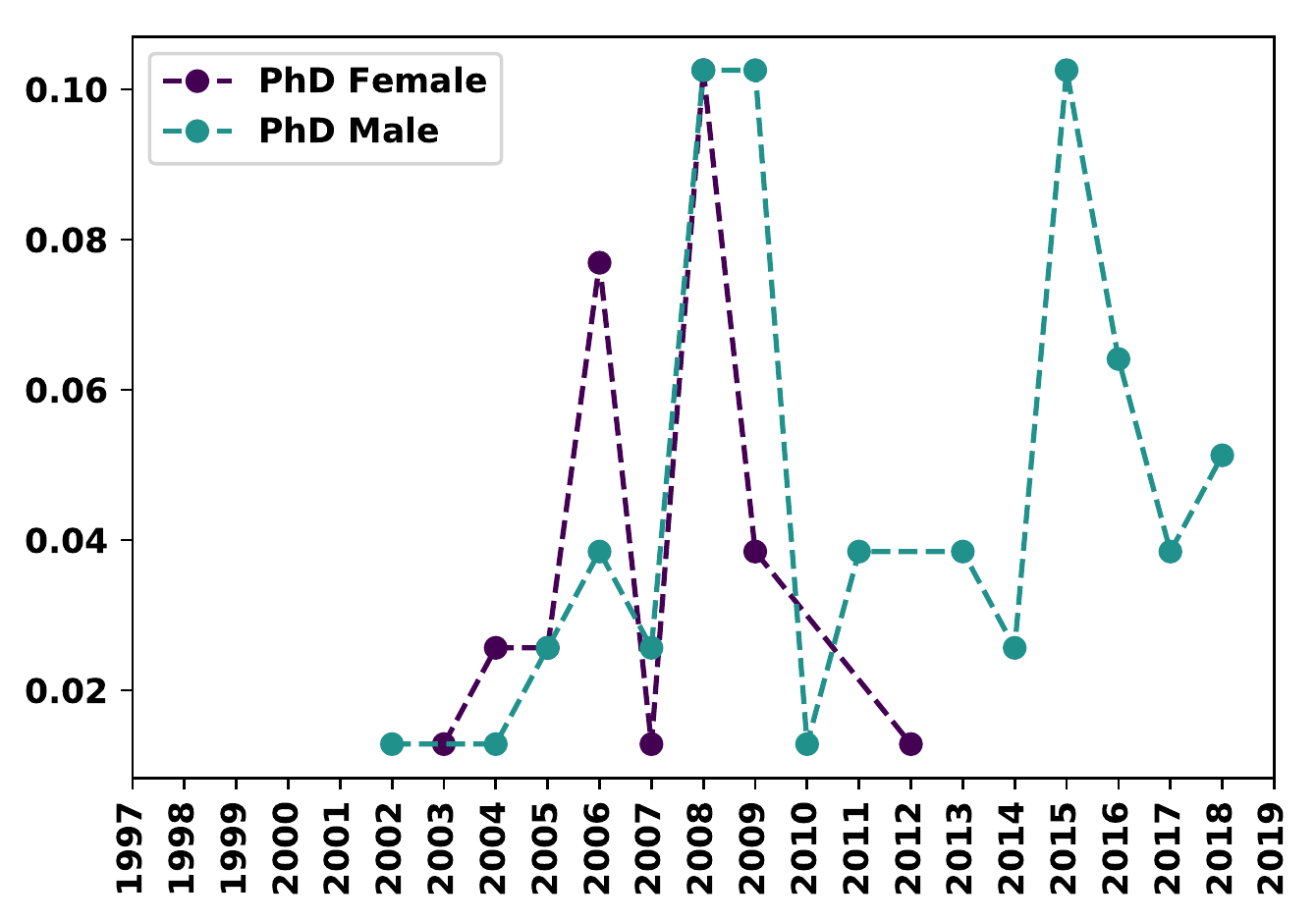}
     }
\caption{Students that completed a degree in a STEM subject at 
UCY, from 1996-2018: (1) BSc; (2) MSc and (3)  PhD.}
\label{Gender_eq_UoC_years}
\end{figure*}

\begin{figure*}[!htbp] 
\subfloat[\label{cypriot_popul_a}]{%
       \includegraphics[width=.48\textwidth, height=0.4\textwidth]{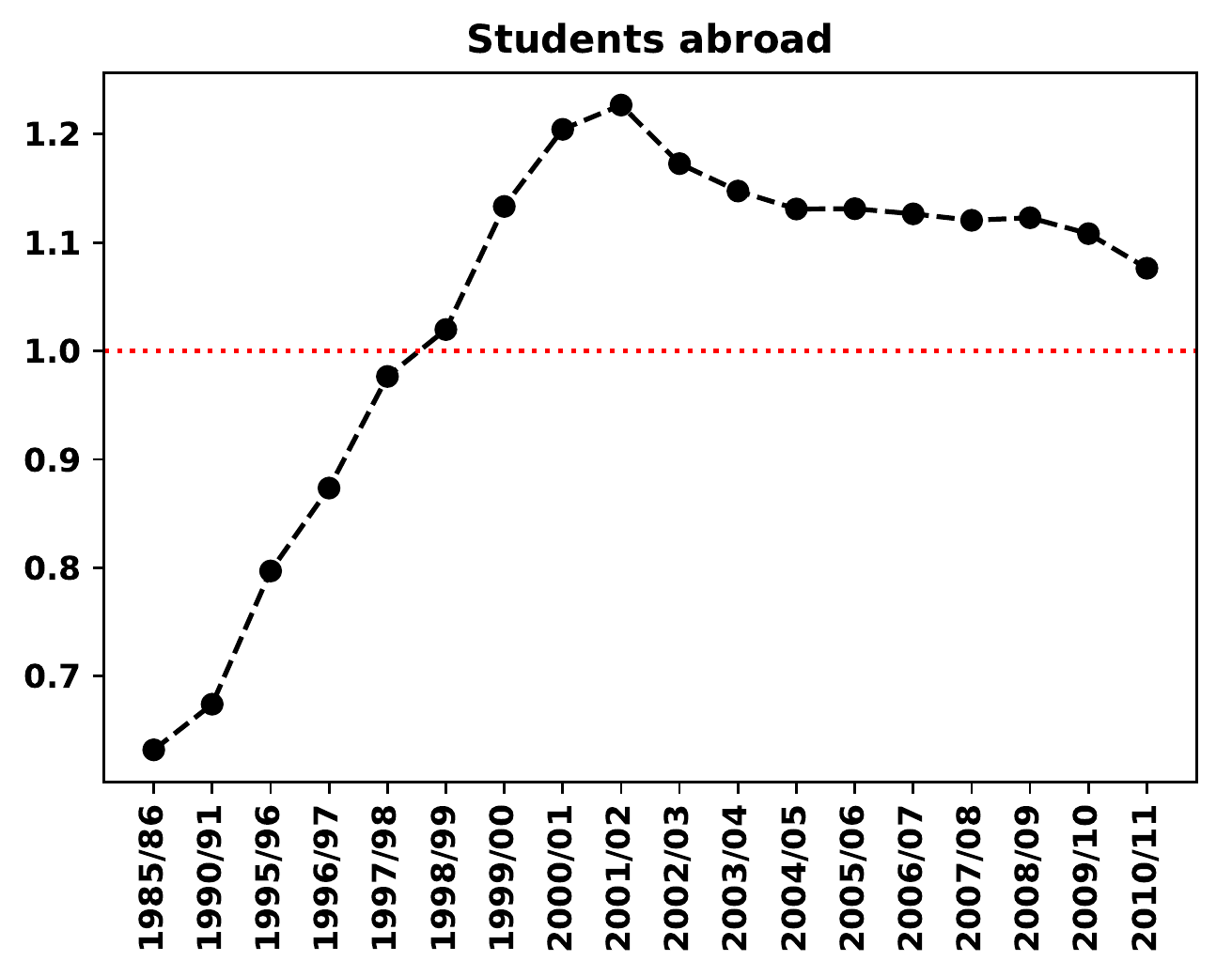}
     }
     \hfill
     \subfloat[\label{cypriot_popul_b}]{%
       \includegraphics[width=.48\textwidth,  height=0.39\textwidth]{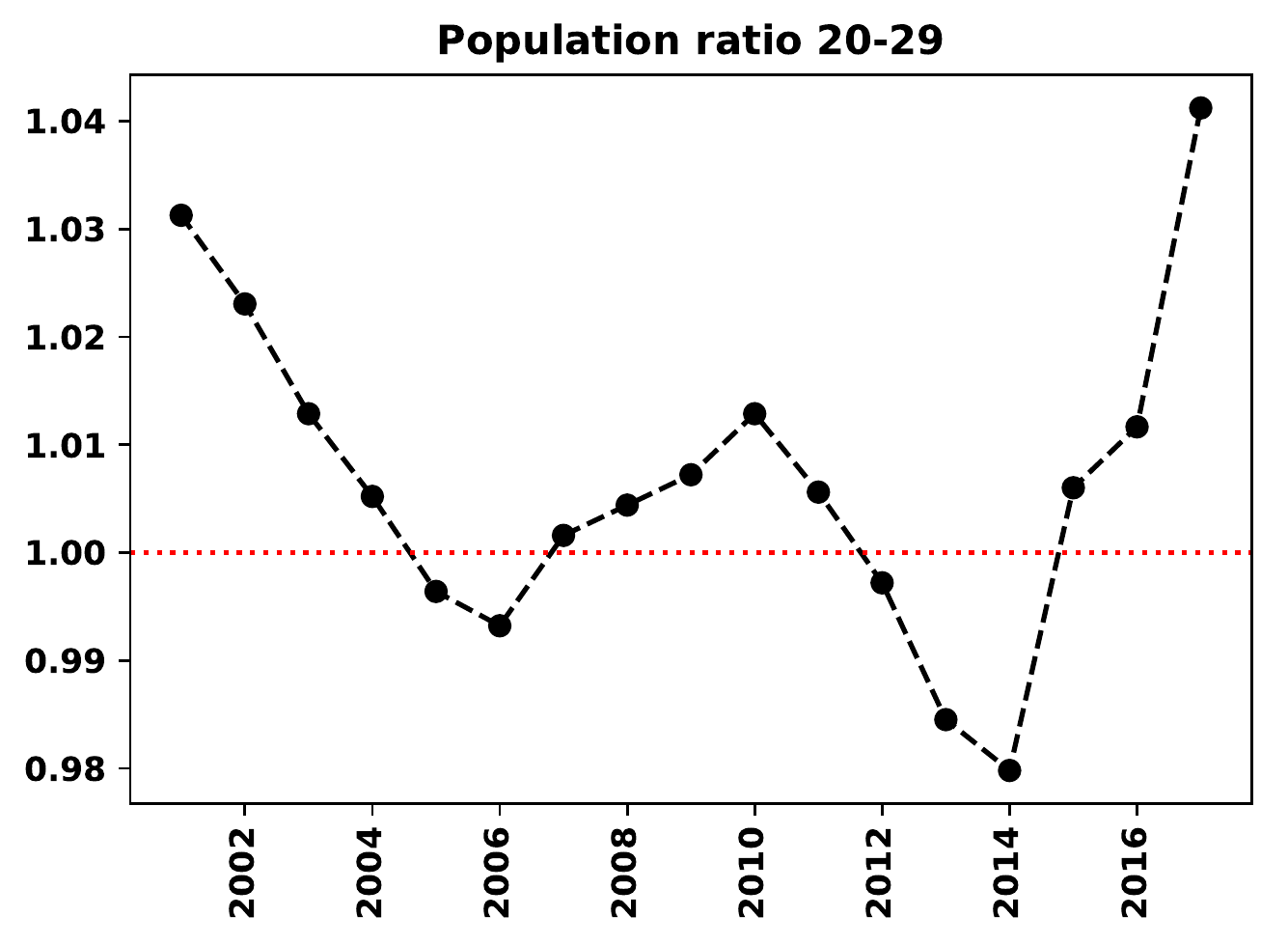}
     }
\caption{Ratio of Cypriot females over Cypriot males over time, for: (1) students studying abroad and (2) population in the age group 20-29.}
\label{cypriot_popul}
\end{figure*}

Analysing the student numbers in STEM subjects in UCY we see that overall that female students are more than male students (55\% versus 45\%, see Table~\ref{tab_grad_perc}).  Histograms of the gender proportions in STEM subjects can be seen in Fig.~\ref{grad_depart_a}, while in Fig.~\ref{grad_depart_b} a gender ratio indicates that, perhaps surprisingly, most departments have more female students than male.  Although this is a positive feature we see in Fig~\ref{degree_per_sex} that most female students do not pursue degrees beyond the BSc, and hence the gender ratio drops significantly at PhD level. This difference between female and male students over the years is illustrated with a line chart in Fig~\ref{Gender_eq_UoC_years}. Again, perhaps surprisingly, we observe that the  proportion of female PhD students has reduced dramatically over the last  ten years\footnote{The university did not provide data for female PhD students after 2012.} From analysing data from the Statistical Service of Cyprus, we find that the
number of female students studying abroad was greater than the 
number of male students, from 1998-2011, although the overall population 
in the age group 20-29 during this period cannot always prove a greater number of women 
versus men. The ratio of female to male students abroad and population 
aged 20-29 is illustrated in Figs.~\ref{cypriot_popul_a} and ~\ref{cypriot_popul_b}, respectively. These data agree with the data from the  UCY showing more women educated in STEM subjects at undergraduate and Masters' level, but still not 
many women entering a research programme afterwards.
\begin{table}
    \begin{center}
      \begin{tabular}{cl|lll|ll}\toprule
           {} & Total & BSc & MSc & PhD & Females &   Males \\ \midrule
           \# GRAD. & 4737 &  3333 & 1326 &  78 &  2605 &  2132 \\ \midrule
          \% GRAD. PER & 100 &  70.35 & 28 &  1.65 &  55 &  45 \\ \bottomrule\\
     \end{tabular}
    \end{center}
     \caption{Number and percentage of students that completed a degree in a STEM subject at UCY, from 1996-2018, per degree and per gender.}
     \label{tab_grad_perc}
\end{table}


\subsubsection{Analysis of STEM faculty data}

\begin{figure}[h!]
    \centering
     \includegraphics[width=.8\textwidth]{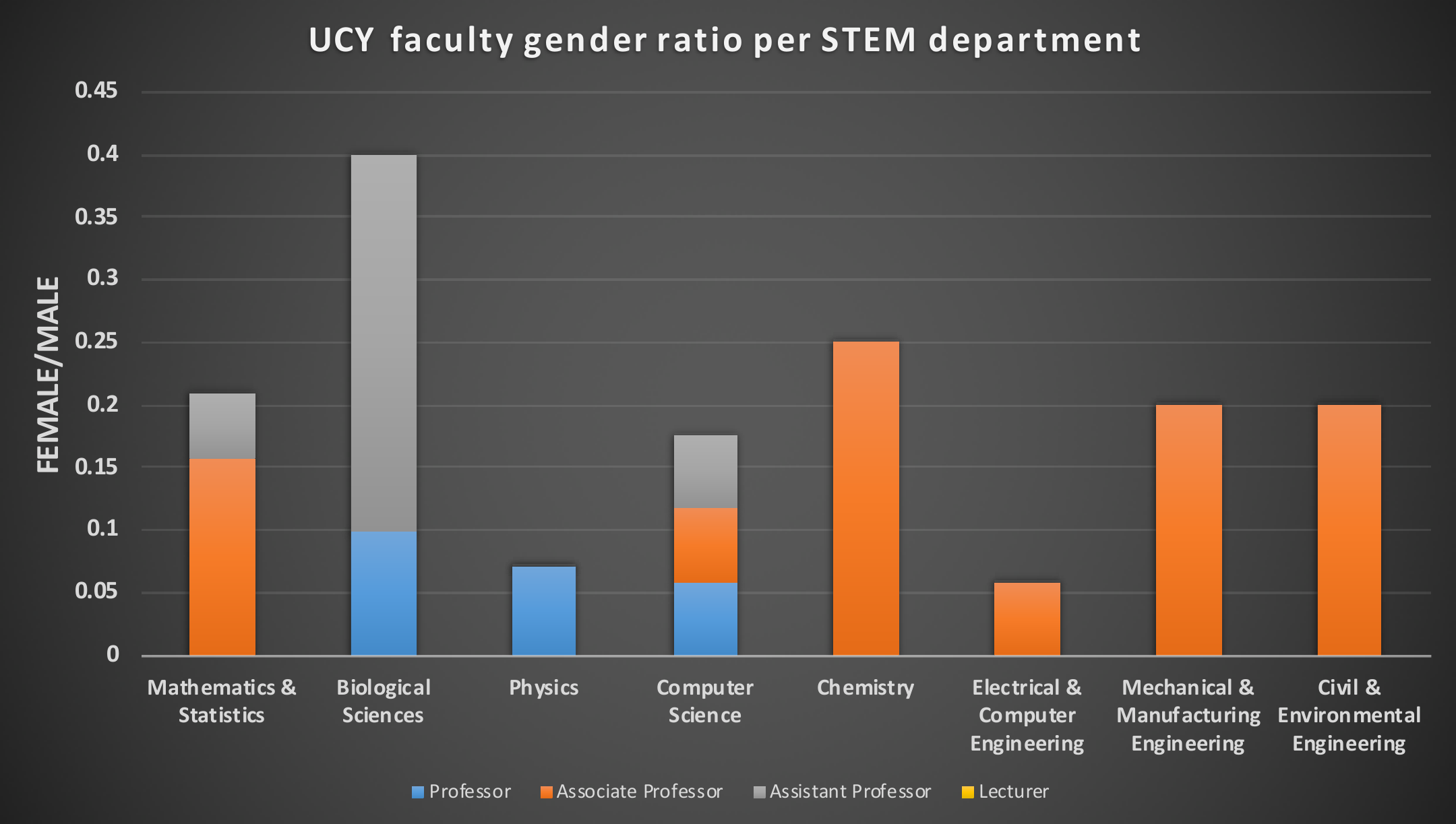}
    \caption{Gender ratio of UCY faculty members in STEM departments.}
    \label{UCY_faculty}
\end{figure}

\begin{figure}[h!]
    \centering
     \includegraphics[width=.8\textwidth]{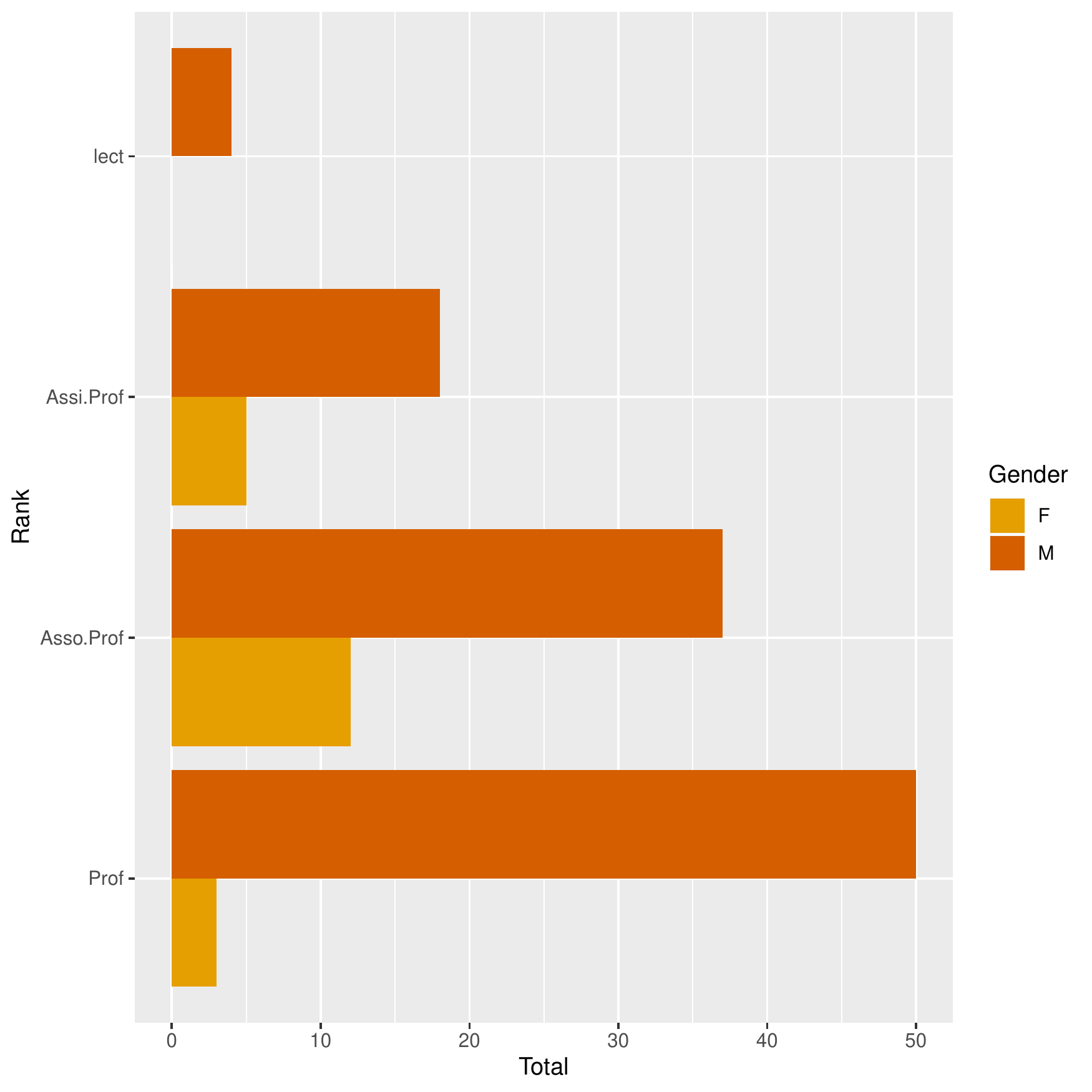}
    \caption{Histogram of the total faculty members for each academic rank in UCY STEM departments.}
    \label{UCY_faculty_summary}
\end{figure}

In contrast to BSc and MSc programs where the number of female students is larger than the number of male students, the female PhD students and faculty members at UCY are much fewer than males, with only 84 female academics in UCY out of the total of 309 in all departments, and 20 out of 129 in STEM departments (see Table~\ref{number_academics}). The higher the academic rank the higher the gender gap with most STEM departments having no female professor (see Figure~\ref{UCY_faculty}). Interestingly, when summarising the overall faculty for each rank in STEM disciplines in UCY, illustrated   Figure~\ref{UCY_faculty_summary}, we observe an overwhelming under-representation of female academics across all ranks with no female lectures and very few female professors. In contrast, for male academics the distribution increases when moving up the ranks with the professors outnumbering the other three ranks. The fact that female academics are not progressing to the professorship level with the same rate as their male colleagues illustrates the phenomenon of the glass ceiling, a widespread phenomenon both in industry and academia whereby female professionals fail to reach the top level of their career path. Further analysis (using a Fisher's exact test in R, p-value=0.01619) indicates that  that there are significant differences across ranks in male to female representation.  Similar trends are observed in many other universities in Europe, with Women representing only the 33\% of the researchers in the European Union,
with a larger gender inequality, i.e. less than 40\% proportion, in the STEM subjects \citep{botella,europcom}. These results suggest that female scientists choose careers outside academia, such as teaching in secondary education, especially if they have a Maths
or Physics degree, as indicated by the larger number of female teachers in high
school compared to male \cite{con}. The challenge of balancing the demands
of a research career with family obligations \cite{kou}, such as child raising
and caring for elderly or sick relatives, in a still somewhat traditional society like Cyprus, probably deters  women from pursuing doctoral
training and academic positions. However, it is still a fact that women are more
educated than men in Cyprus. Eurostat
reported in 2017 a total of 62.1 \% of women in the age group 30-34 that
have successfully completed tertiary education in Cyprus, compared to 43.7 \% of men \citep{eurostat}. Also, women seem to  engage more often than men in lifelong learning and vocational
training.

\begin{table}[h]
  \centering
   \begin{tabular}{cllllll}\toprule
         {}                & Faculty & Females & Males & Ratio \\ \toprule
          All departments  &  309    & 84      &  225  &  0.37 \\ \midrule
          STEM departments &  129    & 20      &  109  &  0.18 \\ \bottomrule \\
   \end{tabular}
    \caption{Number and ratio of female/male of faculty staff in UCY in all and in STEM departments.} 
    \label{number_academics}
\end{table}


\section{Conclusions and recommendations}

Regarding the first part of this report, we developed a new mathematical model to
predict a gender wage gap indicator on the basis of other indicators. This looks promising,
in particular if applied to richer datasets, and possibly to other types of
indicators, than those considered here. Unfortunately, since studies on gender wage gaps are
rather recent, available data are not over a large time period and this is a current limitation of this study. Currently, there is a strong interest in gathering such data and as more data is gathered the applicability of our model will increase.
As it stands now, we split the EU-28 countries in three wage gap levels (bad, medium, best) and found that
\begin{itemize}
\item there is a tendency of countries to remain at their wage gap level
\item ``jumps" of two levels in one year are highly unlikely
\item level 2 countries (medium) have about the same (small) probability to
improve or to get worse.
\end{itemize}
Furthermore, for countries which
start from level 1 (the best level), an improvement in the women/men employment rate and in the women/men higher education is related with a small increase of the probability that the country worsens its situation, passing to level 2.
For countries which start from level 3 (the worst
level), an improvement in the women/men employment rate and in the women/men
higher education is associated with an increase of the probability that the
country improves its situation, passing to level 2, while an
increase in the women/men life expectancy is reducing the probability that the
country passes to level 2.

In the second part of the study, we analysed data provided
by the University of Cyprus and by the Cyprus Statistical Service. The
results seem to indicate that there is a
significant leakage in the pipeline as women go from undergraduate level to
higher stages in academia. To further investigate the reasons for the gender gap in the local community of Cyprus we recommend a comprehensive data analysis in different directions. To achieve this goal we should enhance the existing datasets by detailed information about the employment pathways of the students that completed a degree
 in STEM subjects (either by proposing to the University of Cyprus a lifelong communication with their alumni or extracting data from social networks such as LinkedIn using data mining techniques). Having detailed data we could perform data analysis to examine the careers of male and female students that completed a degree
 in STEM subjects and investigate any social, physiological or psychological reasons why females tend to work in jobs with less-on-average wages than males (e.g. pregnancy, children care, prejudice, less self-confidence, less job-related risky behaviour, etc.). The latter can be implemented in collaboration with researchers from the Department of Sociology and the Department of Psychology of UCY.

Note that the above report is the result of the team's efforts over only one week; no time was available to put these results in their wider  socioeconomic context and seek solutions.  For this, further discussions with AIPFE Cyprus, the University of Cyprus and other interested organisations in Cyprus and in the rest of Europe have been taking or will take place. It is envisioned that eventually this research project will be further developed, leading to as series of tangible recommendations for how to remove barriers for women in science, everywhere. 

\section*{Acknowledgements}
The team acknowledges support from the European Union’s H2020 Research and Innovation Action under Grant Agreement No 741657 \href{https://www.scishops.eu/}{{\color{blue!30!black} (SciShops.eu)}} and from Mathematics for Industry Network (COST Action TD1409), ExxonMobil Cyprus, and the British High Commission in Cyprus.

\bibliographystyle{plain}

\end{document}